\documentclass{aa}  

\usepackage{graphicx}
\usepackage{booktabs}
\usepackage{tablefootnote}
\usepackage{float}
\usepackage{multicol}
\usepackage[dvipsnames]{xcolor}

\usepackage{txfonts}

\usepackage{booktabs}
\usepackage{rotating}
\usepackage[T1]{fontenc}
\usepackage{amsmath}
\usepackage{textgreek}

\usepackage{subcaption}
\usepackage{silence}
\usepackage{natbib}
\bibpunct{(}{)}{;}{a}{}{,}
\usepackage{hyperref} 
\hypersetup{
    colorlinks=true,
    linkcolor=blue,
    citecolor=blue,
    filecolor=blue,
    urlcolor=blue,
}

\begin{document}

\title{Global flow regimes of hot Jupiters}

   \author{
            C. Ak{\i}n\inst{1, 2}
            \and
            K. Heng\inst{2, 3, 4, 5}
            \and
            Jo\~ao M.\ Mendon\c ca \inst{6, 7}
            \and
            R. Deitrick \inst{8, 9}
            \and
            L. Gkouvelis\inst{2}
          }
   \institute{
        Center for Space and Habitability, University of Bern, Gesellschaftsstrasse 6, CH-3012 Bern, Switzerland
        \and
        Universitäts-Sternwarte München, Fakultät für Physik der Ludwig-Maximilians-Universität, Scheinerstraße 1, D-81679 München, Deutschland
        \and 
        ARTORG Center for Biomedical Engineering Research, University of Bern, Murtenstrasse 50, CH-3008, Bern, Switzerland
        \and
        University College London, Department of Physics \& Astronomy, Gower St, London, WC1E 6BT, United Kingdom
        \and
        Astronomy \& Astrophysics Group, Department of Physics, University of Warwick, Coventry CV4 7AL, United Kingdom
        \and
        Department of Physics and Astronomy, University of Southampton, Highfield, Southampton SO17 1BJ, UK
        \and
        School of Ocean and Earth Science, University of Southampton, Southampton, SO14 3ZH, UK
        \and
        Canadian Centre for Climate Modelling and Analysis, Environment and Climate Change Canada, Victoria, BC, Canada
        \and
        School of Earth and Ocean Sciences, University of Victoria, Victoria, BC, Canada
        }
   \date{Received 23 December 2024 / Accepted 16 May 2025}
   
    \abstract{
    The atmospheric dynamics of hot and ultra-hot Jupiters are influenced by the stellar irradiation they receive, which shapes their atmospheric circulation and the underlying wave structures. }
    {We aim to investigate how variations in radiative and dynamical timescales influence global flow regimes, atmospheric circulation efficiency, and the interplay of wave structures across a curated sample of hot Jupiters. In particular, we explore a previously predicted transition in the global flow regime, where enhanced stellar irradiation suppresses the smaller-scale wave and eddy features that feed into superrotating jets and ultimately leads to simpler, day-to-night dominated flows.}
    {We simulate a suite of eight well-studied hot Jupiters with the THOR general circulation model, spanning equilibrium temperatures from about $1100$~K to $2400$~K. We develop a wavelet-based analysis method to decompose simulated wind fields into their underlying wave modes, which we validate on analytical examples. As a preliminary exploration of the flow regime of ultra-hot Jupiters, we perform an additional simulation for WASP-121b, where the mean molecular weight is set to represent an atmosphere dominated by atomic hydrogen.}
    {Our results confirm that increasing stellar irradiation diminishes the efficiency of atmospheric heat redistribution and weakens the contribution of smaller-scale eddy modes critical for sustaining superrotation. As equilibrium temperatures rise, large-scale modes dominate the atmospheric circulation, driving a transition from jet-dominated flows toward day-to-night circulation. Additionally, by artificially lowering the mean molecular weight, we partially restore circulation efficiency and reintroduce a more complex, multi-scale flow pattern. These findings refine our understanding of how atmospheric circulation evolves with increasing irradiation and composition changes, offering a more nuanced framework for interpreting hot and ultra-hot Jupiter atmospheres.}
    {}


   \maketitle

\section{Introduction}
\label{section:Introduction}

Over the past decade and a half, observations by facilities such as the \textit{Spitzer Space Telescope} and the \textit{Hubble Space Telescope} have enabled substantial progress in our understanding of the atmospheric dynamics of short-period, close-in gas giant exoplanets, known as hot Jupiters. These planets, typically residing at orbital distances $<0.1$~AU and expected to be tidally locked to their host stars \citep{guillot_1996, showman_lewis_fortney_2015, dawson_2018}, have become prime laboratories for testing three-dimensional (3D) general circulation models (GCMs). Since the early works of \citet{showman_guillot_2002} and \citet{showman_cooper_2005}, a majority of the GCM efforts have produced atmospheric flows with the following important and ubiquitous features: large-scale eddy and jet structures, significant temperature differences between the day and night sides reaching several hundred Kelvin, and a broad eastward equatorial jet \citep[see][for reviews covering this subject]{heng_showman_2015, showman_tan_parmentier_2020}. 

\begin{figure}
    \centering
    \includegraphics[width=\columnwidth]{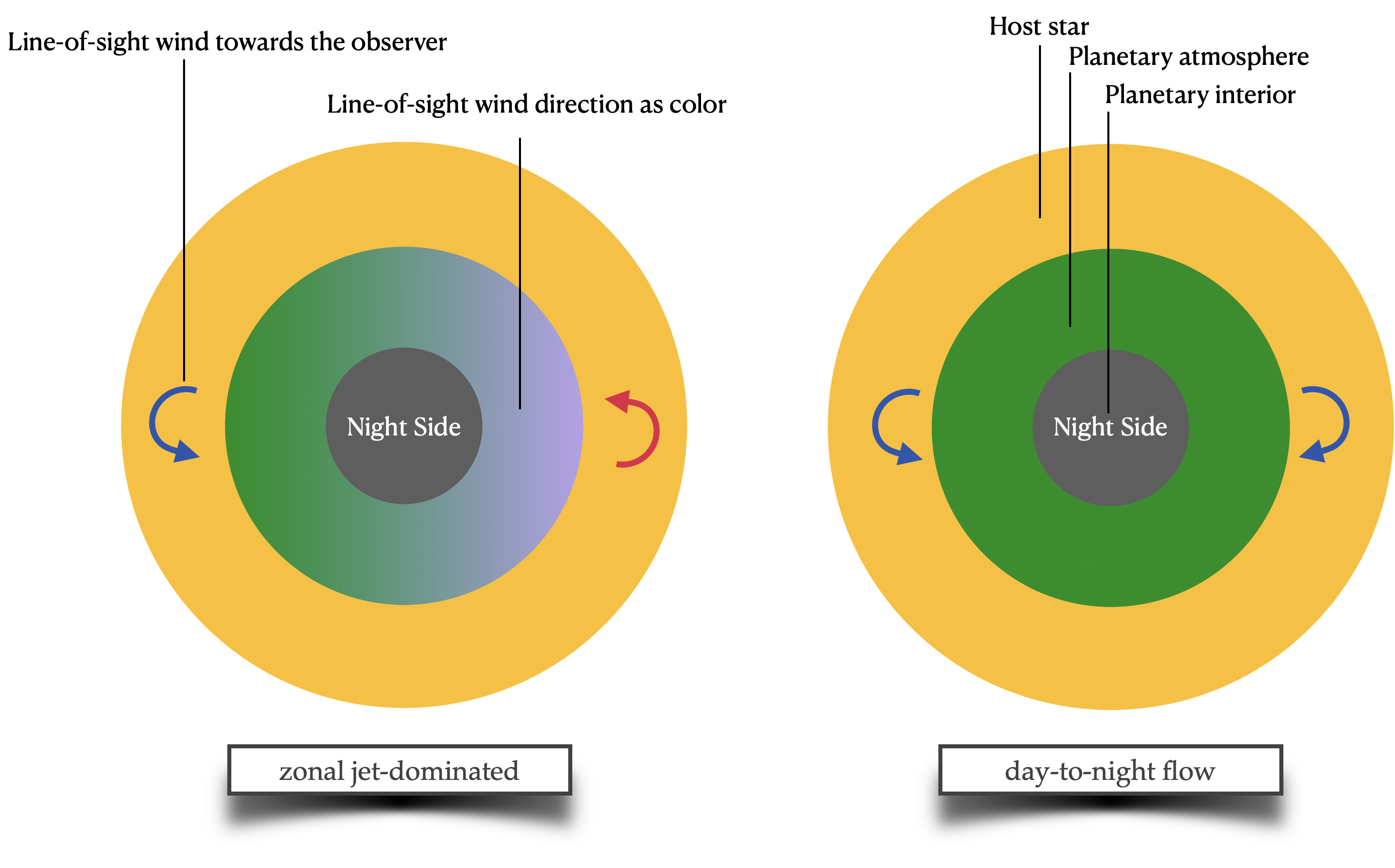}
    \caption{Schematic for the line-of-sight wind maps from Figures \ref{fig:los_terminatorslice_tavg_worotation} and \ref{fig:los_terminatoravg_tavg_worotation_uhj}. The arrow and disk colors are matched such that blue arrows paired with green tones indicate flow moving toward the observer, while red arrows paired with pinkish hues represent flow moving away.}
    \label{fig:visual_aid}
\end{figure}

A key theoretical step in understanding the mechanisms that give rise to these features was provided by \citet{showman_polvani_2011}. They demonstrated that equatorial superrotation can be explained by large-scale, standing planetary waves. These waves are reminiscent of Matsuno-Gill type solutions \citep{matsuno_quasi-geostrophic_1966, gill_simple_1980} to the shallow water equations. They consist primarily of equatorial Kelvin waves and off-equatorial Rossby waves that form as a response to the constant day-night thermal forcing and transport angular momentum toward the equator, thereby maintaining a prograde jet. Building on this, \citet{showman_doppler_2013} expanded this framework by characterizing two distinct atmospheric circulation regimes based on radiative forcing: a jet-dominated regime, where standing Kelvin and Rossby waves sustain equatorial superrotation and a regime dominated by direct day-to-night circulation due to strong damping of these waves. Their study demonstrated how frictional drag and short radiative timescales inhibit jet formation, leading to primarily day-to-night flows. Observationally, these regimes manifest as distinct Doppler signatures, providing a potential method to infer atmospheric wind patterns on hot Jupiters like HD209458b and HD 189733b. Figure~\ref{fig:visual_aid} schematically illustrates the two extremes of this proposed flow transition: a zonal jet-dominated regime and a regime dominated by day-to-night circulation.

Subsequent studies have explored and refined this flow transition framework, using various GCMs and exploring different parts of the parameter space of hot Jupiters. For instance, \citet{mayne_2017} focused on an HD 209458b-like case study, employing the UK Met Office Unified Model (UM) adapted for exoplanetary conditions. Their analysis highlighted that not only waves but also mean-meridional circulations and eddy momentum fluxes are critical for maintaining superrotation. They further emphasized the interplay between vertical and meridional angular momentum transport, showing that while the equatorial jet formation aligns with the wave-driven mechanism outlined by \citet{showman_polvani_2011}, the maintenance of the jet involves more complex interactions.

In contrast, \citet{mendonca_2020} employed Fourier analysis techniques to dissect wave contributions to angular momentum and heat transport in simulations of tidally locked hot Jupiters. Using the UK Met Office UM GCM, \citet{mendonca_2020} isolated the dominant atmospheric wave modes, focusing on stationary and steady components. This methodology revealed that semidiurnal thermal tides, consisting of Kelvin and Rossby waves, drive angular momentum convergence at the equator, forming the strong prograde jet characteristic of these planets. These findings demonstrate how wave activity underpins the observed zonal wind and thermal patterns and further highlights the dependence of such phenomena on planetary rotation rates and radiative timescales.

\citet{tsai_2014} extended the Matsuno-Gill framework by introducing a uniform background zonal wind into their analytical solutions, revealing that such a flow modifies the amplitudes and phase offsets of Kelvin and Rossby waves. They show this using the \citet{dobbs-dixon_agol_2013} simulations for an HD 189733b-like case study and compare their results to those of \citet{showman_polvani_2011}. \citet{hammond_lewis_2021}, using the THOR GCM for HD~189733b and Exo-FMS for a terrestrial planet setup, demonstrated that the atmospheric circulation can be separated into its rotational and divergent components, relating them to different distinct modes of circulation, such as the equatorial jet, standing waves and overturning circulation. Their work showed that while waves drive equatorial jets, overturning circulations significantly contribute to day-night heat transport.

Efforts to generalize Matsuno-Gill solutions to hot Jupiter conditions, such as those by \citet{heng_analytical_2014}, further established that the steady-state wave solutions are sensitive to the strength of radiative forcing, friction, rotation and magnetic fields. These analytical and simplified numerical approaches support the core idea from \citet{showman_doppler_2013} that varying radiative timescales can qualitatively shift the dominant dynamical regime. While the approach of characterizing flow regimes by comparing radiative versus dynamical timescales has been criticized as being oversimplified \citep[e.g.,][]{perez_becker_showman_2013, komacek_showman_2016}, this approach still provides valuable insight into how key parameters influence regime transitions.

Ultra-hot Jupiters, on the other hand, experience intense stellar irradiation and are defined by their daysides reaching sufficiently high temperatures ($T_{\mathrm{eq}} \gtrsim 2500$~K) to dissociate molecular hydrogen into atomic hydrogen \citep[e.g.][]{bell_cowan_2018, arcangeli_2018}. These processes alter radiative timescales and can further modulate the wave patterns and mean-flow structures, potentially extending and complicating the regime transitions described by \citet{showman_doppler_2013}.

Our current work aims to bridge the gap between these theoretical frameworks and the more complex parameter regimes now accessible in GCM studies. We build on the wave analysis formalisms presented in \citet{mendonca_2020}, relaxing periodicity and stationarity assumptions to capture transient and spatially complex wave structures. By systematically exploring a parameter space of hot Jupiters, we examine how wave-mean-flow interactions evolve as we approach the regime of ultra-hot Jupiters. This strategy allows us to test the predictions of theory, including \citet{showman_polvani_2011} and \citet{showman_doppler_2013}, and to refine our understanding of how waves, circulation patterns, and thermochemical processes interact across a wide range of planetary conditions.

\begin{figure*}
    \centering
    \includegraphics[width=\textwidth]{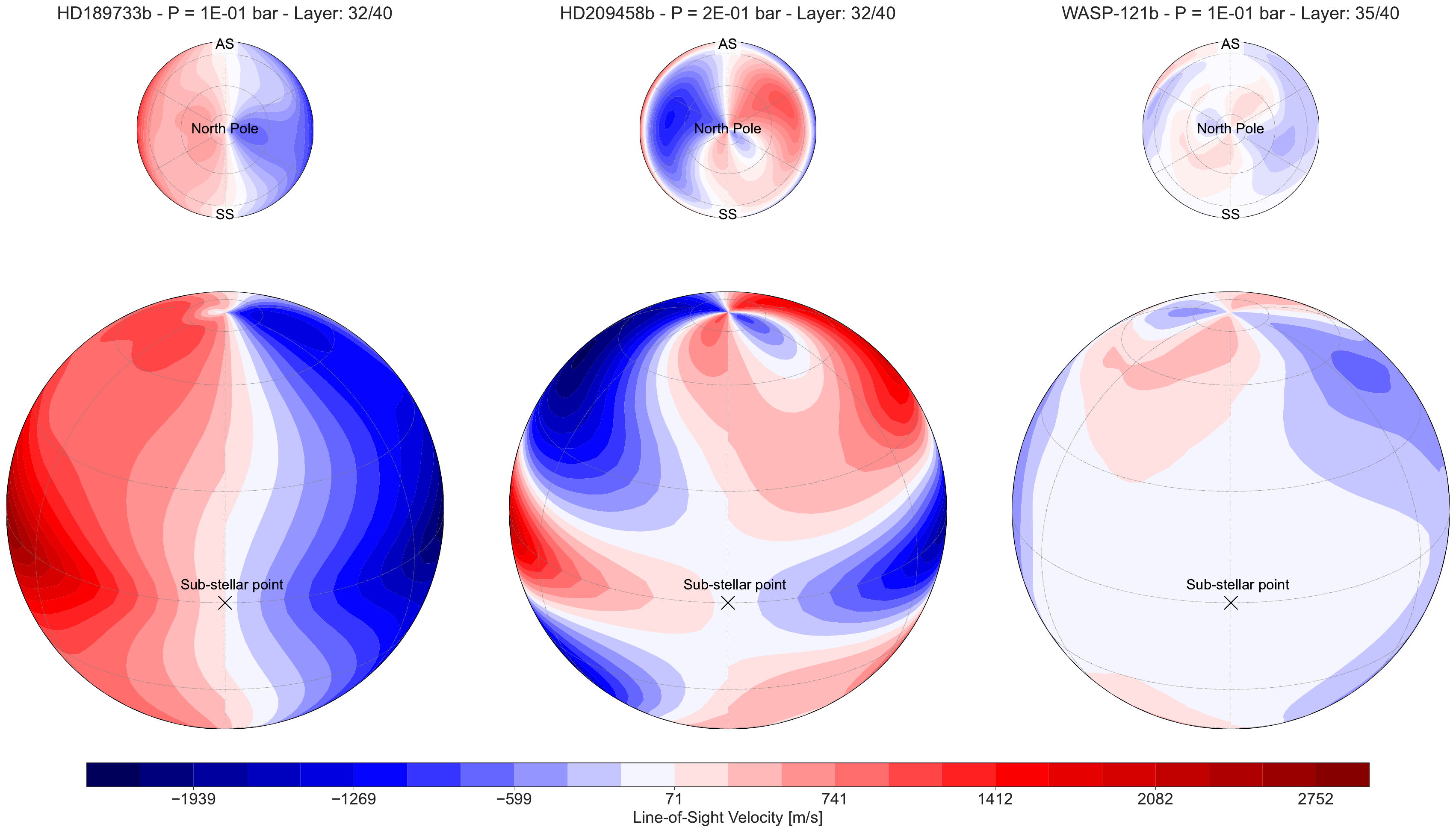}
    \caption{Three-dimensional simulation outputs for three of the eight hot Jupiters in this study. From left to right, the panels showcase (1) a zonal-jet–dominated flow, (2) an intermediate transitional regime, and (3) a day-to-night flow pattern. Each column plots the line-of-sight velocity at the $ P \approx 0.1$ bar atmospheric layer for a single planet. Negative values (blue) denote winds moving toward the observer, while positive values (red) denote winds moving away. The shown results are time-averaged over the last 500 days of their corresponding simulation times.  \textbf{Top}: View from the north pole with the sub-stellar (SS) point and anti-stellar (AS) point marked. \textbf{Bottom}: View from the dayside.}
    \label{fig:showmanetalcomparison}
\end{figure*}

Figure~\ref{fig:showmanetalcomparison} gives a preview of the GCM simulations that we are presenting in the current study. It highlights the three flow regimes witnessed in these simulations, which are driven by different levels of instellation. Illustrated are three representative planets from our sample: one dominated by zonal jets, one dominated by day-to-night (substellar-to-antistellar point) flow, and one in the transitional regime in between. The figure focuses on the upper atmosphere flow at a given isobaric surface. On the left, HD 189733b exemplifies a zonal-jet-dominated regime, as evidenced by the uniformly eastward winds. On the right, WASP-121b displays eastward winds on the evening terminator and westward winds on the morning terminator, creating a pronounced day-to-night flow pattern. The central panel, showing HD 209458b, represents a transitional regime, with a strong equatorial eastward jet coexisting alongside westward flows at higher latitudes.

The remainder of this paper is organized as follows. In Section~\ref{section:THOR}, we present our planet sample and describe our GCM setup using the THOR GCM. Section~\ref{section:Wavelet Analysis} details the wave analysis tools employed in our study. In Section~\ref{section:HJ-to-UHJ}, we review the expected scenarios for flow transitions as we progress toward hotter and more strongly irradiated planets. Section~\ref{section:Results} presents our key findings, including global trends across our sample, the results of our wave analyses, and the implications for hot Jupiter-to-ultra-hot Jupiter transitions. Finally, in Section~\ref{section:Discussion}, we discuss our results in the context of previous studies, highlight the challenges of modeling ultra-hot Jupiters, and suggest potential pathways for future research.
 
\section{Methodology}
\label{section:THOR}
\subsection{Choice of Planet Sequence}
For the systematic exploration of the parameter space, we choose targets that are well-observed and are approximately equally spaced in the irradiation flux they receive from their host star. Our sample consists of HD 189733b, WASP-43b, HD 209458b, WASP-31b, WASP-74b, WASP-19b and WASP-121b. For many of these targets, IR lightcurve observations from \textit{Spitzer Space Telescope} and \textit{Hubble Space Telescope} have already been extensively studied and compared to GCM work, including HD 189733b \citep[e.g.][]{showman_hd189_2009, knutson_hd189_2012, dobbs-dixon_agol_2013, drummond_3d_2018, steinrueck_hd189_2019, flowers_hd189_2019}, HD 209458b \citep[e.g.][]{zellem_hd209_2014, amundsen_hd209_2016, drummond_3d_2018}, WASP-43b \citep[e.g.][]{kataria_atmospheric_2016, mendonca_w43b_2018b, deitrick_thor_2022}, WASP-31b \citep{kataria_atmospheric_2016}, WASP-19b \citep{wong_wasp19b_2016}, WASP-121b \citep{parmentier_w121b_2018, lee_w121b_2022}. Additionally, the planets HD 189733b, WASP-43b, and HD 209458b serve as benchmark objects from the \textit{Hubble Space Telescope} and \textit{Spitzer Space Telescope} eras. A comprehensive list of planetary parameters is given in Table \ref{tab:planet_data}. For all the presented runs, a 1x solar metallicity was assumed rather than adopted from the referenced papers.

\begin{table*}
\centering
\setlength{\tabcolsep}{2.5pt}
\renewcommand{\arraystretch}{1.3}
\caption{Parameter values assumed for the GCM simulations. \protect\footnotemark}
\begin{tabular}{@{}lccccccccccc@{}}
\toprule
\textbf{Planet} & $\mathbf{R}$\newline[$R_{J}$] & $\mathbf{\Omega}$\newline[rad/s] & $\mathbf{g_P}$\newline[m/s$^2$] & $\mathbf{c_P}$\newline[J/(kg K)] & $\mathbf{T_{eff}}$\newline[K] & $\mathbf{T_{\star}}$\newline[K] & $\mathbf{a}$\newline[au] & $\mathbf{R_{\star}}$\newline[$R_\odot$] & $\mathbf{z_{top}}$\newline[H] & $\mathbf{t_{run}}$\newline[days] & \textbf{Ref.} \\
\midrule
HD 189733b & 1.216 & 3.28e-05 & 21.40 & 12383 & 1117 & 4875 & 0.03297 & 0.805  & 19.43 & 5000 & 1, 2 \\
WASP-43b  & 0.930 & 8.94e-05 & 47.00 & 12383 & 1379 & 4400 & 0.01420 & 0.600  & 20.67 & 4000 & 3 \\
HD 209458b & 1.390 & 2.06e-05 & 9.18  & 12383 & 1455 & 6092 & 0.04634 & 1.190  & 21.39 & 4000 & 1, 2 \\
WASP-31b  & 1.549 & 2.14e-05 & 4.60  & 12383 & 1575 & 6302 & 0.04659 & 1.252  & 16.37 & 4000 & 4 \\
WASP-17b  & 1.932 & 1.95e-05 & 31.60  & 12383 & 1755 & 6550 & 0.05125 & 1.583  & 18.27 & 4000 & 4 \\
WASP-74b  & 1.560 & 3.40e-05 & 8.90  & 12383 & 1917 & 5970 & 0.03700 & 1.640  & 19.71 & 4000 & 5 \\
WASP-19b  & 1.395 & 9.22e-05 & 14.20  & 12383 & 2067 & 5440 & 0.01616 & 1.004  & 17.44 & 4000 & 6 \\
WASP-121b & 1.807 & 5.70e-05 & 9.40  & 12383 & 2358 & 6460 & 0.02544 & 1.458  & 16.89 & 4000 & 7 \\
\bottomrule
\end{tabular}
\label{tab:planet_data}
\end{table*}

\subsection{The Dynamical Core}
In our study, we employ the THOR GCM to simulate atmospheric circulation on exoplanets, as introduced and developed in a series of papers by \citet{mendonca_thor_2016}, \citet{mendonca_w43b_2018a},  \citet{mendonca_w43b_2018b},  \citet{deitrick_thor2.0_2020}, \citet{deitrick_thor_2022}, and \cite{noti_examining_2023}. THOR is an open-source, fully three-dimensional dynamical core that solves the non-hydrostatic deep (NHD) Euler equations using a horizontally explicit, vertically implicit \citep{tomita_satoh_2002, wicker_skamarock_2002} time integrator on an icosahedral grid \citep{tomita_shallow_2001}. This split-time integration scheme enables THOR to take a reasonable time step while still maintaining numerical stability with vertically propagating waves in its solutions—waves that would otherwise impose a much smaller time step under the Courant–Friedrichs–Lewy condition. On the other hand, our choice of grid structure overcomes the common limitations of latitude-longitude grids \citep{staniforth_thuburn_2012}, which suffer from singularities and resolution issues at the poles. This issue extends to all non-uniform grid structures characterized by fixed points with significant variations in cell sizes or crowding. The icosahedral grid as introduced in \citet{tomita_shallow_2001} and \citet{tomita_new_2004} incorporates a quasi-uniform horizontal grid structure. The THOR model uses an altitude grid in the vertical direction and is equipped with two possible configurations. The first setup, as introduced and explained in \citet{mendonca_thor_2016} and \citet{deitrick_thor2.0_2020}, utilizes a uniformly spaced altitude grid. The bottom layer altitude is calculated from the ideal gas law, and the top of the atmosphere is a free parameter set by the user. A new option introduced in \citet{noti_examining_2023} allows finer control of the vertical spacing, incorporating a non-uniform vertical structure. Although the default THOR model primarily solves the NHD Euler equations, as detailed in \citet{deitrick_thor2.0_2020}, \citet{deitrick_thor_2022} and \citet{noti_examining_2023}, it also offers the flexibility to incorporate various levels of simplification into the governing equations. These include options such as the quasi-hydrostatic deep and the hydrostatic shallow sets of equations, each with their respective assumptions. For this study, our focus remains on the NHD equations, with no simplifying assumptions applied.

All of our newly run models assume solar metallicity and are cloud-free. The comparison cases presented in Section \ref{section:Discussion} are run with the same setup as presented in 
\citet{deitrick_thor_2022}. All 8 of our main model runs have been completed with $g_{level} = 5$ corresponding to a $\approx 2^{\circ}$ horizontal resolution and 40 vertical layers, uniformly spaced in altitude space with the bottom boundary pressure corresponding to $P_{ref} = 1000$ bar. The physical height of the uppermost vertical layer varies between models due to their varying sizes and numerical considerations (explained in Subsection \ref{subsec:stability_and_numerics}). This results in a range of pressure levels at the top of the atmosphere, ranging from $\approx 10^{-4}$ bar for our coldest planet in the sample HD 189733b  to $\approx 0.05$ bar for the hottest planet in our sample WASP-121b. These values represent the minimum global average pressure, indicating that the entire planet extends to this level in pressure space. However, the actual minimum pressure on the night side reaches $\approx 10^{-6}$ bar across all simulations.

\footnotetext{The unit [H] refers to pressure scale heights as defined in Eq.~\ref{eq:scale_height}. \textbf{References}: 1-\citet{torres2008}, 2-\citet{boyajian2015}, 3-\citet{hellier2011}, 4-\citet{anderson2011}, 5-\citet{stassun2017}, 6-\citet{cartes-zuleta2020}, 7-\citet{delrez2016}.}

\subsection{Stability and Numerical Parameters}
\label{subsec:stability_and_numerics}

At present, all state-of-the-art general circulation models require a numerical dissipation scheme \citep{Jablonowski2011, heng_2011a}, representing the dissipation of energy at small, unresolved scales. The simulation domain, especially for giant planets like the ones explored in this study, has characteristic length scales so large that they result in minimum resolved lengths that are still multiple orders of magnitude larger than molecular scales. This limitation usually results in a grid-scale energy build-up. To this end, the THOR model incorporates horizontal and vertical hyperdiffusion terms \citep{mendonca_thor_2016, deitrick_thor2.0_2020} and a divergence damping operator. These are, respectively, higher-order gradient terms and a Laplacian that acts on the divergence field. The chosen order of these operators determines on which scales the dissipation schemes selectively act on, with higher orders indicating smaller effective scales. Lastly, a sponge layer \citep{Jablonowski2011, mendonca_w43b_2018b} can be introduced upwards of a certain altitude to filter out unphysical vertical waves that get reflected from the top of the atmosphere. The sponge layer has the form of a Rayleigh friction term and damps the zonal, meridional and vertical wind velocities to the zonal averages. All of these additions are modular and can be customized to ensure the stability of a given model run. 

Importantly, the choice of dissipation scheme can influence not only numerical stability but also the emergent large-scale circulation. Recent studies using the BOB model have shown that even brief increases in small-scale energy loss can steer the flow toward markedly different large-scale states \citep{skinner_cho_2022, skinner_cho_2025}. This indicates that while dissipation schemes are primarily introduced for stability, their design can non-trivially shape the physical interpretation of a simulation. Consequently, models that maintain stability through low spatial resolution, strong explicit viscosity, or basal drag risk suppressing the very small-scale motions that play a key role in shaping global circulation and temperature patterns \citep{skinner_cho_2021}. \citet{hammond_numerical_2022} have performed an extensive review of the effect these dissipation schemes have on the THOR GCM and the reader is advised to consult that work for a comprehensive explanation of these methods.

For the models presented in this work, we utilize all three dissipation schemes. We use 6th-order horizontal and 4th-order vertical hyperdiffusion terms, together with a 6th-order divergence damping operator. Our exact parameters for the numerical setup can be found in Table \ref{tab:run_parameters}. Our approach mirrors the recent studies done with the THOR GCM \citep{deitrick_thor_2022, noti_examining_2023}, aiming for comparability.

In addition to the numerical dissipation schemes discussed in this section, many other GCMs utilize an interior boundary drag formulation to account for unmodelled physical processes such as vertical turbulent mixing \citep{li_goodman_2010}, Lorentz-force braking \citep{perna_2010}, or other processes. THOR does include a Rayleigh drag scheme for the bottom of the simulation domain \citep{deitrick_thor2.0_2020}, but we do not employ it in this work. This option of bottom Rayleigh drag is independent of the top-of-atmosphere sponge layer described above. It is worth emphasizing that we are exploring simulations with a deep lower boundary ($\sim$1000 bar). The atmospheric processes in the upper atmosphere drive these exoplanet atmospheres, and since it would be prohibitively computationally expensive to reach a statistically steady state in the deepest layers due to the very large thermal inertia, we set the lower boundary free of drag, which, in any case, is observationally unconstrained. While there is qualitative motivation for including these dissipation schemes, they are, strictly speaking unphysical and are not specified from first principles. In other words, they are numerical tools manifesting as additional terms in the Navier-Stokes equation.

\begin{table}
\centering
\small
\renewcommand{\arraystretch}{1.2}
\caption{Values of hyper-parameters}
\label{tab:run_parameters}
\resizebox{\columnwidth}{!}{%
\begin{tabular}{@{}l c l r@{}}
\toprule
\textbf{Symbol} & \textbf{Descriptor} & \textbf{Unit} & \textbf{Value} \\ 
\midrule
$\Delta t$ & Time step & [s] & 300 \\
$g_{\mathrm{level}}$ & Grid refinement level & [-] & 5 \\ 
$v_{\mathrm{level}}$ & Number of vertical levels & [-] & 40 \\
$D_{\mathrm{hyp, h}}$ & Horizontal 6\textsuperscript{th}-order hyperdiffusion coefficient & [-] & 0.0025 \\ 
$D_{\mathrm{hyp, v}}$ & Vertical 6\textsuperscript{th}-order hyperdiffusion coefficient & [-] & 0.001 \\
$D_{\mathrm{div}}$ & 4\textsuperscript{th}-order divergence damping coefficient & [-] & 0.01 \\
$k_{\mathrm{s}p}$ & Sponge layer strength (Horizontal/Vertical) & [s$^{-1}$] & $10^{-3}$/$10^{-4}$ \\
$\eta_{\mathrm{sp}}$ & Bottom of sponge layer (fraction of $z_{top}$) & [-] & 0.8 \\
$\eta_{\mathrm{lats}}$ & Number of latitude bins used in sponge layer & [-] & 20 \\
\bottomrule
\end{tabular}
}
\end{table}

\subsection{Radiative Transfer}
For the radiative transfer (RT) component of our simulations, THOR is equipped with three RT schemes: The radiative transfer component in our model includes three key modules. The model incorporates a two-stream double-gray radiative transfer module, as outlined in \citet{mendonca_thor_2016},  \citet{mendonca_w43b_2018b} and \citet{deitrick_thor2.0_2020}. In addition, we employ a two-stream "picket-fence" non-gray radiative transfer module, as introduced by \citet{noti_examining_2023} and modeled on the approach by \citet{lee_simulating_2021}. Furthermore, the model includes an "improved two-stream" method for multi-wavelength radiative transfer, named Alfrodull (or THOR+HELIOS in \citealt{deitrick_thor_2022}), which provides an accurate treatment of scattering by medium-sized and large aerosols. This method is originally described in \citet{heng_analytical_2017} and \citet{heng_analytical_2018}, and involves coupling the THOR GCM with the HELIOS radiative transfer code, as detailed in \citet{deitrick_thor_2022} and referenced in \citet{malik_helios_2017, malik_self-luminous_2019}. 

For the main results of this study, we utilize the picket-fence scheme. This decision is grounded in the findings of \cite{noti_examining_2023}, which highlighted the picket-fence scheme's optimal balance between computational efficiency and improved physical accuracy. As discussed in Section \ref{section:Introduction}, our goal is not to conduct a detailed study of each object in our hot Jupiter sample, but rather to perform a parameter-space-wide analysis of dynamical behavior. While the double-gray method provides a simple and fast calculation suitable for parameter-space explorations, the reality of these planets is that gaseous radiative transfer should be non-gray \citep{showman_tan_parmentier_2020}. Although the improved two-stream method would offer a more accurate representation, it requires significantly longer computation times. Thus, the picket-fence scheme offers an excellent trade-off.

The picket-fence method \citet{1935MNRAS..96...21C}, \citet{parmentier_picket_fence_2014} and \citet{parmentier_picket_fence_2015} is a non-gray analytical radiative-transfer approach designed to model the radiative processes of giant planets more accurately than the double-gray approximation. In contrast to the double-gray model, where two constant opacities, $\kappa_{\text{lw}}$ and $\kappa_{\text{sw}}$, are used, the picket-fence model employs opacities that are locally interpolated from a pressure–temperature grid. The infrared (IR) opacities are represented by the Rosseland mean opacities, as calculated by \citet{freedman_2014} for a wide range of gaseous giant planets, which we adopt for our study.

The picket-fence model further refines the conventional visual and infrared bands used in double-gray approximation by dividing them into five distinct channels: three for the visual band, representing stellar insolation, and two for the IR band, representing planetary emission. This division into multiple channels aims to capture both continuum and atomic/molecular line opacities. Using the locally interpolated IR opacities, the opacity structures for the additional IR and visual bands are derived through analytical relationships as detailed in \citet{parmentier_picket_fence_2014}. For a more comprehensive explanation of the method, we direct the reader to the implementation paper \citep{noti_examining_2023}, which adheres to the conventions established by \citet{lee_simulating_2021}.

The specific setup for our suite of hot Jupiters assumes solar metallicity for all of our sample planets. We additionally adopt the Rosseland mean opacity tables that include contributions from TiO and VO (as they are presented in \citet{freedman_2014}). Since our hot Jupiter sample covers a large range of equilibrium temperatures, including the border region between the hot Jupiter and ultra-hot Jupiter classifications, the opacity contributions of these metal oxide species become relevant. 

\subsection{Characteristic Flow Quantities}
\label{sec:Characteristic Flow Quantities}
The following sections will discuss flow properties and wave analyses of objects across the parameter space of hot Jupiter dynamics. It is therefore of importance that we introduce commonly defined flow properties \citep[e.g.][]{showman_guillot_2002, perna_effects_2012, kataria_atmospheric_2016, showman_tan_parmentier_2020}, proposing them as a natural language through which we can compare our planet sample. To begin with, we calculate the pressure scale height defined as 

\begin{equation}
\label{eq:scale_height}
    H = \frac{k_{B}T}{m g}
\end{equation}
where $k_{B}$ [J/K] is the Boltzmann constant, $T$ [K] the temperature of the gas,  $g$ [m/s\(^2\)] the gravity, $m$ [kg/mol] the molar mass of the gas. Next, we calculate the Rossby number \citep{holton_1992, showman_guillot_2002, menou_2003}, defined as
\begin{equation}
    Ro = \frac{U}{f L}
\end{equation}
which quantifies the relative strength of atmospheric dynamics compared to Coriolis effects induced by the planet's rotation. A Rossby number of order unity indicates a transitional regime where both advection and Coriolis forces are significant, which is crucial for understanding the atmospheric circulation patterns of hot and ultra-hot Jupiters. Here $U$ [m/s] is the characteristic wind speed, $L$ [m] the characteristic length scale, and $f = 2 \Omega \sin{\text{\texttheta}}$ [s\(^{-1}\)] the Coriolis parameter with $\Omega$ [rad/s] being the rotation rate of the planet and \texttheta [rad] representing the latitude. In our study, we adopt the root mean square (RMS) velocity as the characteristic wind speed. Following the convention of \cite{noti_examining_2023}, we adopt the Rossby deformation radius $L_{D}$ \citep{gill_1982, showman_guillot_2002}, defined as, 

\begin{equation}
    L_{D} = \frac{N D}{f}
\end{equation}

as the characteristic length scale. This choice is motivated by $L_{D}$'s ability to approximate the length scale over which gravity waves and rotational effects balance, thereby providing a meaningful measure for categorizing atmospheric dynamics across our parameter space. Here $N$ [s\(^{-1}\)] is the Brunt–Väisälä frequency and $D$ [m] is the characteristic vertical length scale of a given atmosphere. We choose $D=H$ following our previously mentioned convention. The Rossby deformation radius gives us an approximate length scale over which processes like gravity waves become comparable to the rotational effects and is thus a natural choice for the previously defined characteristic horizontal length scale $L$. 

The Brunt–Väisälä frequency, which is a measure of the stability of a stratified fluid to vertical displacements, is defined as \citep{emanuel1994atmospheric}
\begin{equation}
    N = \sqrt{\frac{g}{\theta} \frac{d\theta}{dz}}
\end{equation}
where $\theta$ is the potential temperature. The potential temperature is a measure of the temperature that is corrected for adiabatic expansion. Under the assumption of a constant specific heat capacity $c_{P}$, which is valid for our GCM runs (see Section \ref{section:HJ-to-UHJ}), the potential temperature $\theta$ and specific entropy $S$ are are related by $S = c_{P} \ln{\theta}$. Consequently, the potential temperature $\theta$, serves as a proxy for entropy, facilitating our analysis of atmospheric stability and dynamics, as further discussed in Section \ref{section:Discussion} .

In this study, we investigate the interplay between stellar irradiation and atmospheric dynamics, focusing on how radiative and dynamical timescales influence atmospheric circulation and wave structures. Specifically, we calculate the associated radiative ($t_{\text{rad}}$) and dynamical 
($t_{\text{adv}}$) timescales in our simulations \citep{showman_guillot_2002}: 

\begin{align}
t_{adv} &\sim \frac{R_{p}}{u_{max}} \\ 
t_{rad} &\sim \frac{c_{P} P }{\sigma_{SB} g T^{3}}
\end{align}
where $R_{p}$ is the planetary radius, $u_{max}$ is the maximum zonal wind velocity, $c_{P}$ is the specific heat capacity, $\sigma_{SB}$ the Stefan-Boltzmann constant and $T$ the temperature. These timescales provide us with a simple yet effective way of analyzing the efficiency of heat circulation and the formation of atmospheric wave structures, such as equatorial superrotation. For colder hot Jupiters, $\left(t_{\text{adv}}\right)$ in the upper atmosphere is shorter or comparable to $\left(t_{\text{rad}}\right)$, allowing dynamical processes to shape the atmospheric flow. Conversely, for hotter objects, the radiative timescale becomes so short that dynamical processes are less effective in forming large-scale structures, particularly in the upper atmosphere. 

\subsection{Wave Analysis}
\label{section:Wavelet Analysis}
To decompose and analyze wave structures assumed to exist within the highly non-linear solutions produced by a GCM, we require a method well-suited for signals that vary across space (non-stationary) and over time (transient). Some authors have tackled the problem of analyzing wave behavior through Fourier analysis \citep{mendonca_2020}, using a \citet{wheeler_convectively_1999}-style analysis \citep{tan_atmospheric_2021}, or employing a Helmholtz decomposition \citep{hammond_lewis_2021, lewis_hammond_2022}. While Fourier analysis is computationally efficient, easy to understand, and generalizable to multiple dimensions—providing information about the involved spatial scales—it does not offer insights into transient behavior and the time-varying interplay between dynamical structures. \citet{wheeler_convectively_1999}-type analyses incorporate a Fourier decomposition of transient structures by averaging out a background assumed to be constant, delivering insights into wave behavior. However, they can be challenging to reproduce exactly due to the variety of possible data preprocessing steps (e.g., the length of the time-averaging window or whether additional filtering techniques are applied), and typically require overlaid theoretical solutions to "guide the eye". Helmholtz decompositions are adaptable methods that split the field into rotational and divergent components, providing insights into the physical interplay between, for example, Rossby and Kelvin waves. However, they do not offer information regarding the characteristic length and time scales of these waves.

We employ the continuous wavelet transformation (CWT) \citep{farge_wavelet_1992} in this work because it allows us to analyze non-stationary and transient structures while providing dominant length and time scales of the physical processes involved from an agnostic perspective to assumptions about the magnitude of these quantities. The only input required is the range of spatial scales over which the data is decomposed; selecting a sufficiently large range effectively mitigates concerns about this parameter. Additionally, wavelet analysis enables us to recover full dispersion relations $\omega\left( k\right)$, where $\omega$ is the temporal frequency of the wave and $k$ is the wavenumber of the wave. While \citet{wheeler_convectively_1999}-style analyses also provide dispersion relations, they cannot simultaneously offer the spatial distribution of decomposition coefficients that help identify which underlying waves are present. Instead, they often overplot theoretical dispersion relations for inertio-gravity, Kelvin, and Rossby waves, which are not fitted to the data nor physically justified since they stem from linear analysis.

Wavelet analysis is a time-frequency method that enhances the Fourier transform's capabilities by capturing both time and frequency (or space and spatial frequency) details from a signal. Using wavelets as decomposition functions distributes the precision of the Fourier transform between time and frequency, allowing analysis of transient structures without the stationarity assumption required by Fourier analysis. However, the information recovered remains constrained by the Heisenberg-Gabor uncertainty \footnote{Also known as the Gabor limit \citep{gabor_1945}, after Dennis Gabor, this notion extends Heisenberg’s relation to signal analysis; in essence, a function and its Fourier transform cannot both be strictly bounded.} principle \citep{morlet_1982}, resulting in a natural spread within the derived dispersion relations for wave structures. While optimization techniques \citep{moca_time-frequency_2021} can reduce this spread, they are beyond the scope of this study.

\begin{figure}
    \centering
    \includegraphics[width=\columnwidth]{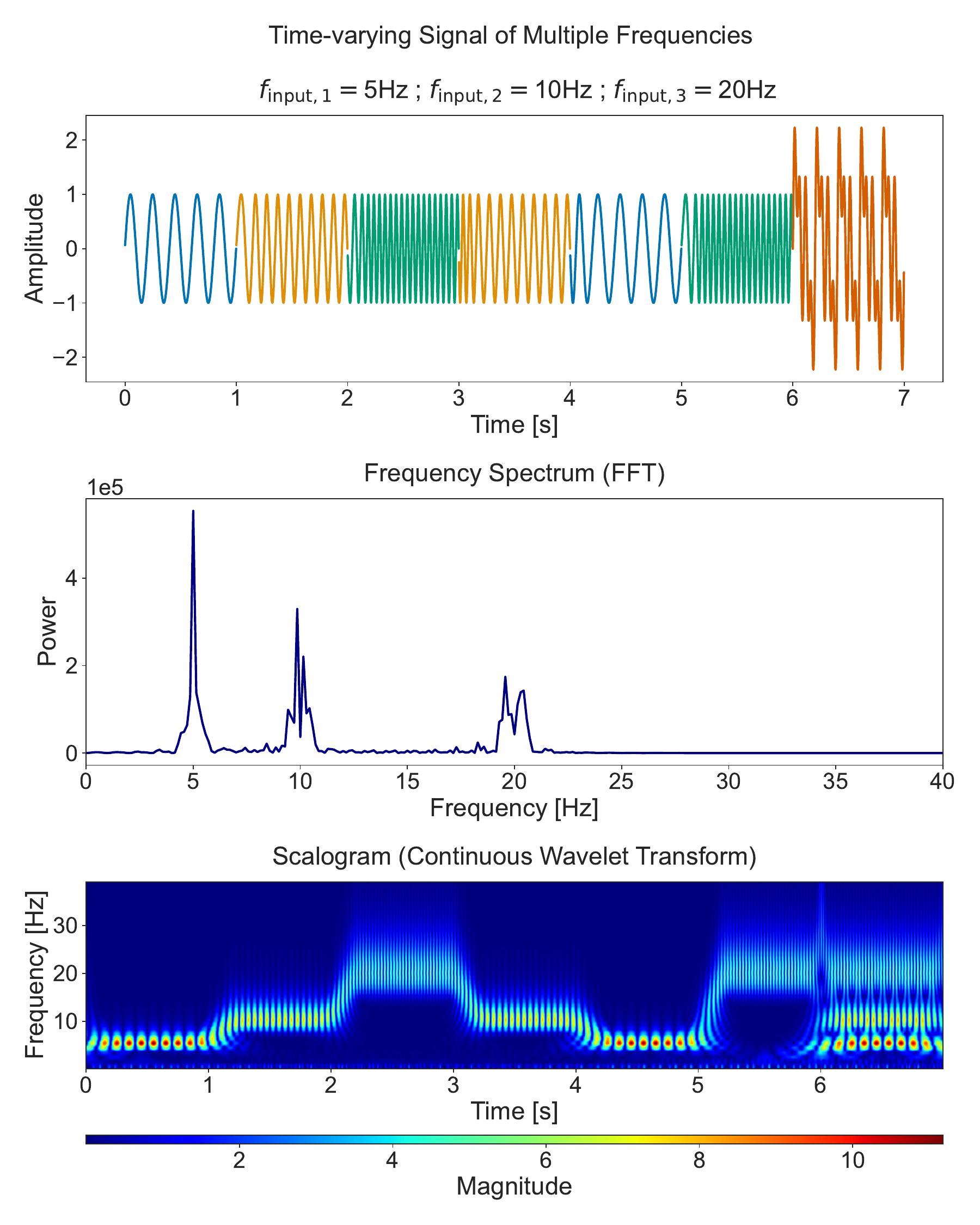}
    \caption{A simple comparison illustrating the differences between a Fourier transform (middle panel) and a continuous wavelet transform (CWT) (bottom panel). \textbf{Top}: A simple time-varying sinusoidal signal composed of successive pure sine waves at different frequencies, culminating in the final second where all three frequencies overlap. Signals of the same frequency have the same color. \textbf{Middle}: The fast Fourier transform (FFT) power spectrum of the signal, providing a single, time-averaged frequency-domain view. \textbf{Bottom}: A CWT scalogram of the same signal, showing how the power at each frequency component changes over time.} 
    \label{fig:fft_vs_wavelet}
\end{figure}

A simple illustration of the differences between a Fourier decomposition and wavelet decomposition can be found in Figure \ref{fig:fft_vs_wavelet}. For illustration, we construct a toy model that consists of an arbitrary series of sinusoidal waves of three different frequencies. The original signal displayed on the top panel consists of three pure sine waves of a given amplitude. The time series evolves as we move from a sine wave with low frequency to sine waves of higher frequencies, and lastly, we combine all three signals. As is evident from the illustration, the Fourier decomposition of such a signal would only be able to tell you that there are three distinct sine waves within this signal. Simply judging from the Fourier transform, we would not be able to differentiate between a sequential version of the same time series, for example the last part with all three signals combined would have a nearly identical Fourier space representation just by itself. Contrastingly, the bottom panel showing the wavelet decomposition result tells us the time-dependent story of how these simple waves contribute to the overall signal. The above-mentioned spread in the recovered frequency values is also visible.

Before describing our methodology for recovering the wave structures, we would like to define what wavelet functions are and the continuous wavelet transformation we are using, in addition to our choice of a wavelet function, defining the natural language of this decomposition. A wavelet function describes a short-term oscillation and is defined \citep{farge_wavelet_1992} to have zero mean and is confined in both spatial (physical) and frequency (Fourier) domains. We adopt the definition given in \citet{torrence_practical_1998} and define the CWT as 
\begin{equation}
\label{eq:CWT}
    W_{n}\left(s\right) = \sum_{\tilde{n}=0}^{N-1} x_{\tilde{n}} \Psi^{*} \left[ \frac{\left(\tilde{n} - n\right) \delta t}{s}\right],
\end{equation}
where $x_{n}$ are the $n$ points in a discrete time series equally spaced out in time by $\delta t$, $s$ is the non-dimensional wavelet scale that scales the size of that wavelet function and $\Psi^{*}$ describes the complex conjugate of the wavelet function. The key components of a CWT are the dimensionless scale and time parameters. Given a time-dependent signal, considering the example of the zonal (east-west) component of the GCM velocity field, these parameters stretch the scanning wavelet function and translate it across the dimensionless time axis to calculate the "match" between the zonal wind components and the used wavelet function at a given scale. For our wavelet function, we adopt a Morlet \citep{morlet_1982} wavelet for the decomposition. The Morlet wavelet function in its general form is defined as, 
\begin{equation}
\label{eq:morlet}
    \Psi_{0}\left(\eta\right) = \pi^{-1/4} e^{i\omega_{0}\eta} e^{-\eta^{2}/2}
\end{equation} 
a plane wave that is modulated by a Gaussian, where $\eta$ is the dimensionless time parameter and $\omega_{0}$ defines the non-dimensional central frequency of the wavelet. Expressed in simple terms, Morlet wavelets are a family of functions describing normalized and decaying local oscillations. Our selection of this wavelet function is grounded in both practical considerations, namely, its successful application in previous atmospheric science studies \citep{torrence_practical_1998}, and theoretical justification, specifically its capacity to maintain an optimal balance in the time-frequency, or in our case space and spatial frequency, trade-off governed by the Heisenberg–Gabor uncertainty \citep{morlet_1982}. As an extension of the Gabor wavelet \citep{gabor_1945}, the Morlet wavelet was specifically designed to minimize this uncertainty.

We utilize the already existing \textit{pyWavelets} Python package \citep{lee_pywaveletspywt_2023} for the wavelet decomposition due to its computational efficiency together with the fact that their implementation also follows the \citet{torrence_practical_1998} conventions.

As previously mentioned, the only fixed input parameter is a range of spatial scales to decompose our input signal. Following our definitions, these input scales are dimensionless and define how our input wavelet functions will be "stretched". Given that the sampling period $P_{\text{sampling}}$ of the input signal is known, one can relate these dimensionless quantities to their physical values. For recovering spatial frequencies, the sampling period $P_{\text{sampling}}$ corresponds to the size of a grid cell. Alternatively, if one is interested in recovering temporal frequencies, this would be the computational time-step. To be able to interpret our results in a physically meaningful way, we choose to express our input scales as zonal wavenumbers calculated by
\begin{equation}
    k = \frac{2 \pi \omega_0}{s P_{\text{sampling}}},
\end{equation}
where $\omega_{0}$ is the dimensionless central frequency of the wavelet as given in Eq. \ref{eq:morlet}, $s$ the dimensionless scale parameter as given in Eq. \ref{eq:CWT}, and $P_{\text{sampling}}$ the above-mentioned sampling period of a given signal. To make the discussion uniform across our hot Jupiter sample, we further adopt the following normalizations
\begin{equation}
    k_{\text{char}} = \frac{2 \pi}{2 \pi R_p} = \frac{1}{R_p} \\
    \omega_{\text{char}} = \Omega_p
\end{equation}
\begin{equation}
    \hat{k} = \frac{k}{k_{\text{char}}} = k R_p \\
    \hat{\omega} = \frac{\omega}{\omega_{\text{char}}} = \frac{\omega}{\Omega_p}
\end{equation}
where $\omega_{\text{char}}$ is the characteristic temporal frequency, $k_{\text{char}}$ is the characteristic zonal wavenumber, $R_p$ is the radius and $\Omega_p$ the rotation rate of the exoplanet. The choice of the characteristic wavenumber $k_{\text{char}}$ is motivated by the desire to express wavenumbers in a dimensionless form that directly relates to a planet’s size. If we consider a wave pattern that completes one full cycle around the planet at the equator, its wavelength must equal the planetary circumference $2 \pi R_{p}$. By choosing $k_{\text{char}} = \frac{1}{R_p}$, we define a dimensionless wavenumber $\hat{k}$ in which $\hat{k} = 1$ represents a wave that fits exactly once around the planet’s circumference. Larger or smaller values of $\hat{k}$ then naturally correspond to waves that repeat multiple times or fractionally around the planet. This normalization thus provides an intuitive measure of how the spatial scales of atmospheric structures relate to the size of the planet itself. Our input wavenumbers are identical for all the presented analyses and correspond to the normalized wavenumber range $\hat{k} = \left[ 0.1, 10\right]$.

Using the methodology described above, we recover a set of coefficients for the wavelet functions defined for each input scale and for each point in time. In other words, we recover a time-dependent power spectrum of our input zonal wavenumbers. As such, we can identify the dominant modes of circulation and recover their dispersion relations $\omega \left( k \right)$, for a given temporal frequency $\omega$ and zonal wavenumber $k$.

To address another important point, we would like to mention how the temporal evolution of the GCM solutions plays a role in our analysis. Conventionally, the GCM results shown in exoplanetary literature are a time average of what would be considered the steady state result after the spin-up period is discarded. This convention relies on the fact that the intention is to study the average global climate state, rather than the temporal evolution of small-scale features. In this sense, the temporal evolution does not portray the true development of how the steady state came into existence. However, in our case we are interested in the time-dependent production and dissipation of waves, because these are the mechanism by which the global climate state (the equatorial jet and the related day-night contrast) is formed, and thus the physical evolution and development of the flow pattern, as the GCM is "spun up" from rest, contains relevant information.

\subsection{The Transition from Hot Jupiters to Ultra-Hot Jupiters}
\label{section:HJ-to-UHJ}
For our fiducial runs, we assume a standard $H_{2}$-dominated atmosphere. This choice implies an implicit assumption on the mean molecular weight that modifies the specific heat capacity $c_{p}$ and the specific gas constant $R_{d}$. While this assumption proves reasonable for standard hot Jupiter cases, ultra-hot Jupiters pose a different challenge. As mentioned in \citet{bell_cowan_2018} and \citet{tan_atmospheric_2019}, around the $\approx 2200-2500$ K point for equilibrium temperatures we move to a different chemical regime where molecular hydrogen $\rm H_2$ starts getting dissociated on the day side of the atmosphere. 

As we probe the lower end of this temperature range with WASP-121b, with an equilibrium temperature of $T_{eq} = 2358$ K, we attempt to mimic this change in behavior with a simple, idealized case. We assume $\chi_{\rm H} = 1$ and $m_{\rm H} \approx u$, resulting in an idealized mean molecular weight $\mu_{m} = 1$. This situation corresponds to an atmosphere where hydrogen exists purely in its atomic form. Note that this is different from what was proposed by  \citet{bell_cowan_2018}, who suggested that the dayside of an ultra-hot Jupiter is dominated by atomic hydrogen, while its nightside is dominated by molecular hydrogen.  
. As mentioned above, this case is meant to illustrate the upper boundary of the proposed mechanism for modifying the flow. The results are presented in Section \ref{section:Discussion}.

\section{Results}
\label{section:Results}
\subsection{Atmospheric Structure of GCM Sequence}
To establish the correspondence between previous parametric exploration studies such as \citet{perna_effects_2012}, \citet{showman_polvani_2011} and \citet{noti_examining_2023}, together with studies focusing on a sample of observed exoplanets such as \citet{showman_doppler_2013} and \citet{kataria_atmospheric_2016}, we would like to go over the general circulation properties of our planet sample. As many previous studies have indicated (e.g. \citealt{perna_effects_2012, bell_cowan_2018, jones_2022}), the relationship between the irradiation temperature and the heat circulation patterns and efficiencies go through several distinct transitions as we increase/decrease in irradiation temperature. 

\begin{figure*}[ht]
    \centering
    \includegraphics[width=\textwidth]{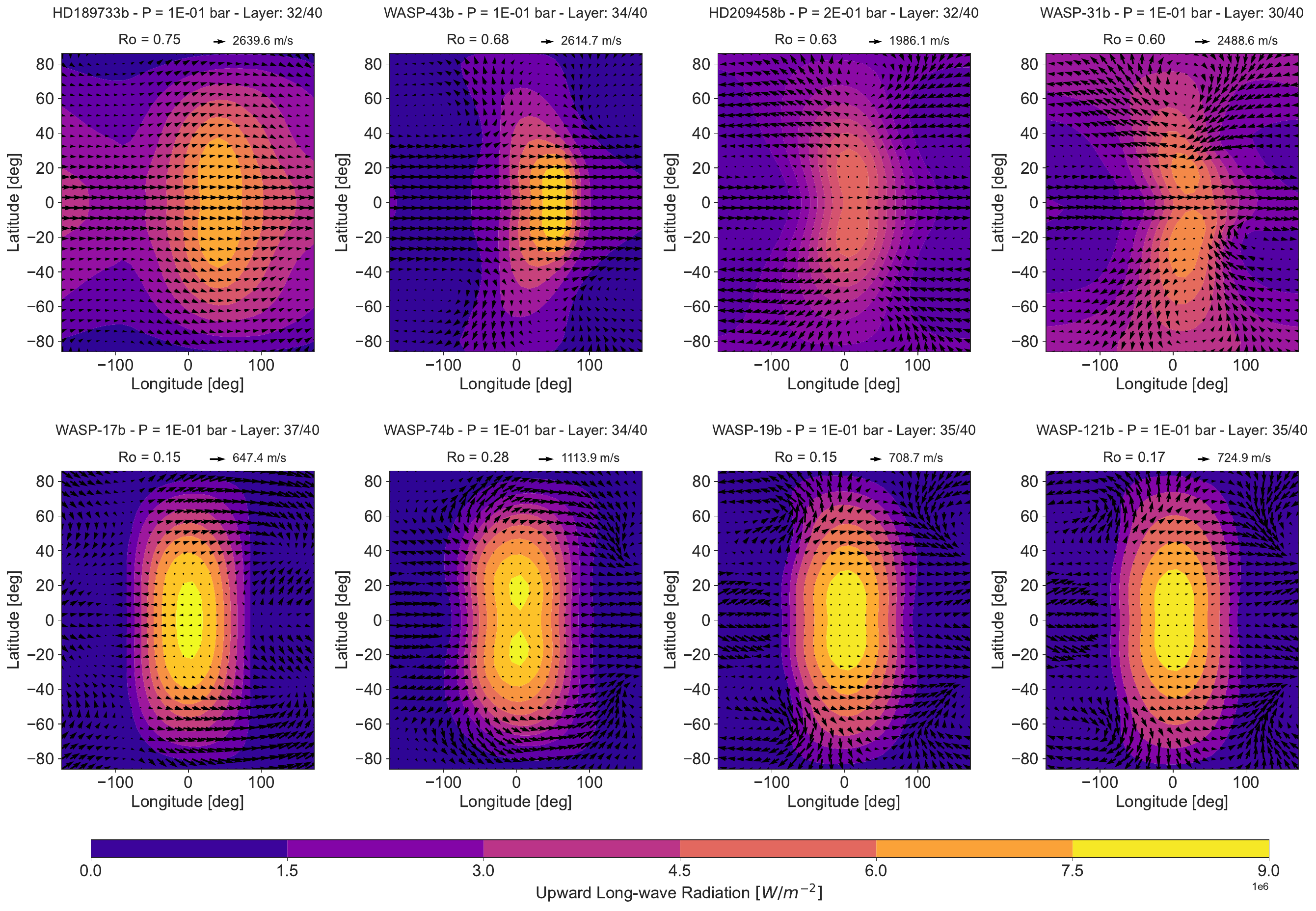}
    \caption{Upward long-wave radiation plots (coloured contours) overlaid with the wind field (arrows) at the pressure surface $P = 0.1$ bar. Plots are time-averaged over the last 500 days of their corresponding simulation times.}
    \label{fig:OLRulev0.1}
\end{figure*}

\begin{figure*}
    \centering
    \includegraphics[width=\textwidth]{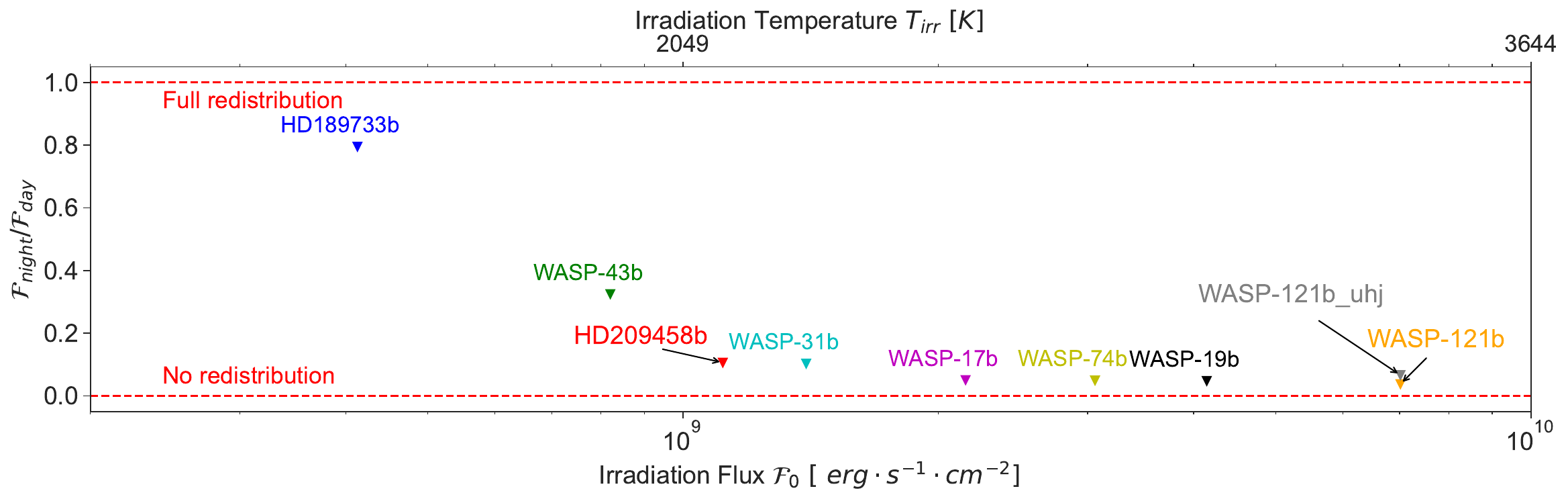}
    \caption{Heat circulation efficiency, which is defined as the ratio of the nightside to dayside upward long-wave radiation fluxes, as a function of the stellar flux received by the hot Jupiter.}
    \label{fig:perna_plot}
\end{figure*}

In Figure \ref{fig:OLRulev0.1}, we show the upward long-wave radiation flux patterns overlaid with winds at the approximate IR photosphere ($P \approx 0.1$ bar). As equilibrium temperatures rise, going from the upper left corner panel to the bottom right corner panel in Figure \ref{fig:OLRulev0.1}, there is a notable increase in the day-to-night upward long-wave radiation contrast accompanied by a gradual reduction in the eastward hotspot offset. For the colder planets, in the upper row, a distinctive Matsuno-Gill \citep{matsuno_quasi-geostrophic_1966, gill_simple_1980} chevron pattern emerges. With higher equilibrium temperatures (or increased irradiation temperature), this pattern gradually diminishes, evolving into a singular, prominent hotspot and reinforcing the increasingly pronounced day–night contrast observed in the upward long-wave radiation maps. However, to quantify this contrast more systematically, we examine the night-to-day flux ratio. By examining this ratio, we can evaluate how effectively the global circulation transports heat away from the sub-stellar point, thereby refining our understanding of the evolving flow patterns suggested by the maps in Figure \ref{fig:OLRulev0.1}. In Figure \ref{fig:perna_plot}, we plot the night-to-day upward long-wave radiation flux ratio for each planet and observe a decreasing trend with increasing instellation. This declining ratio serves as a proxy for reduced heat circulation efficiency, confirming that circulation efficiency decreases with higher instellation until the ultra-hot Jupiter regime ($\approx 2500~\mathrm{K}$ is reached. Our study only extends until the beginning of the ultra-hot Jupiter regime with our mock ultra-hot Jupiter example WASP-121b which shows a marginal increase in the heat circulation efficiency. These results are consistent with previous studies \citep[e.g.,][]{jones_2022, bell_cowan_2018} and reproduce the anticipated observational and theoretical behavior. It is important that we again underline the caveat that for the slight increase in our "mock ultra-hot Jupiter" case, we focus on changing mean molecular weight without taking into account the subsequent heating/cooling; this approach is intended solely as a qualitative reproduction of the expected behavior rather than as an extensive comparison to previous studies.

\begin{figure*}
    \centering
    \includegraphics[width=\textwidth]{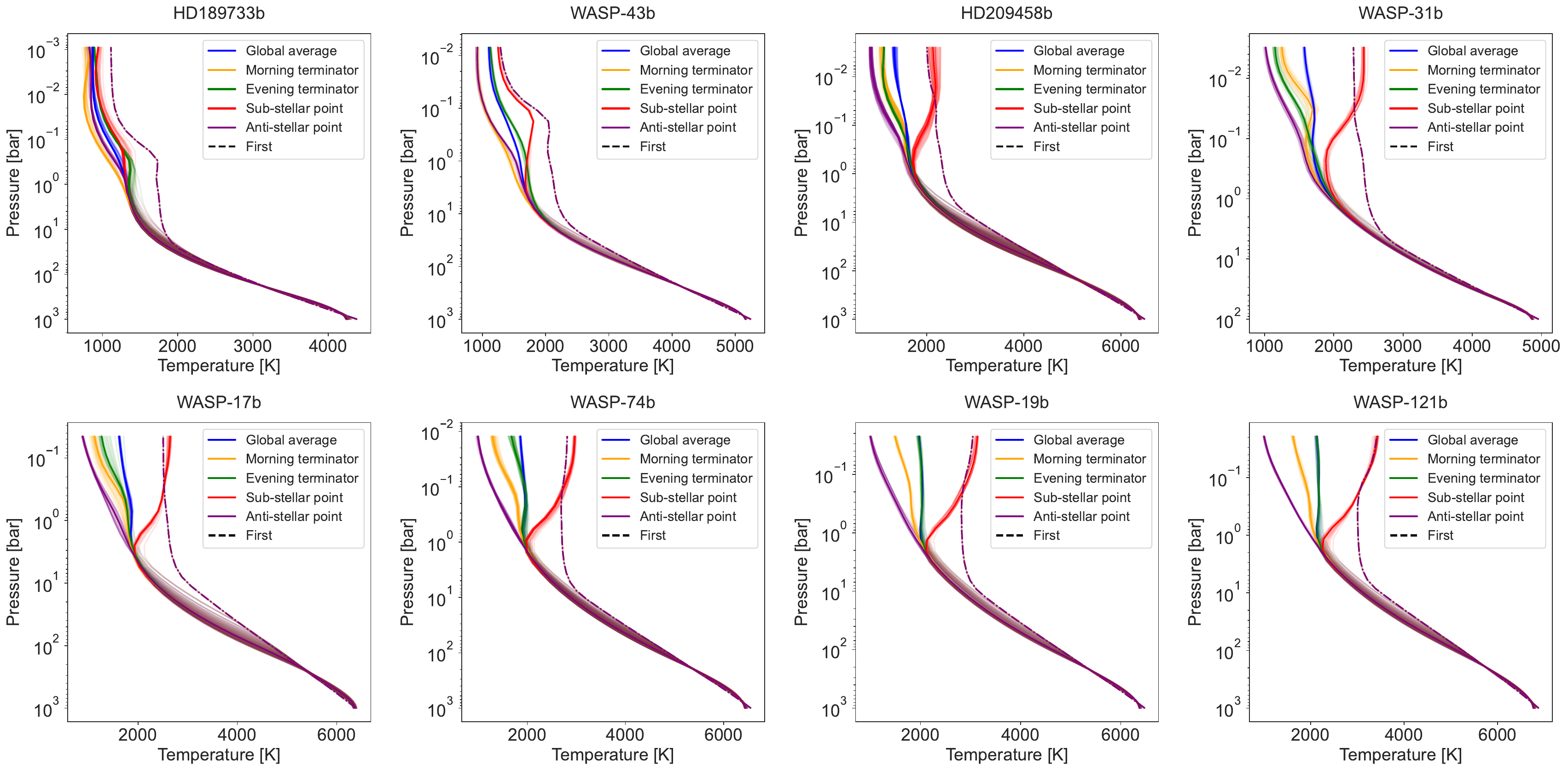}
    \caption{$T$-$P$ profiles from simulations of the 8 hot Jupiters in our GCM sample, including that of the morning terminator, evening terminator, substellar point and antistellar point.  Also shown is the globally averaged profile.  The initial temperature-pressure profile is shown as a dot-dashed curve.  Each curve is accompanied by a shaded region that shows the temporal evolution of each profile, where the first 500 days of spin-up have been excluded.}
    \label{fig:TP_tavg_init_final}
\end{figure*}

Figure \ref{fig:TP_tavg_init_final} illustrates the $T$-$P$ profiles for each planet. The initial profile, which is globally identical, is superimposed with the time-averaged profiles, demonstrating that initial conditions are effectively forgotten. The global $T$-$P$ structure evolves into a profile that is adiabatic in the interior and nearly isothermal in the upper atmosphere. Significant temperature contrasts arise between the sub-stellar and anti-stellar points for every planet, growing from a few hundred Kelvin for the coldest planet in the upper-left corner to a couple of thousand Kelvin for the hottest planet in the bottom-right corner. The temperature contrast between the morning and evening terminators also exhibits a distinctive evolution. The contrast between the terminators initially narrows (e.g., HD 209458b, WASP-31b, WASP-17b) and then widens again in hotter planets, ultimately stabilizing at a difference of a few hundred Kelvin despite the widening gap between the sub-stellar and anti-stellar points. We discuss the implications of this shift for the global flow regime in Section \ref{sec:Global Flow Transition}. A thermal inversion layer located between the $P \approx 0.1 - 1~\mathrm{bar}$  atmospheric layers is evident in most sub-stellar point $T$-$P$ profiles, becoming more pronounced for hotter planets.

\begin{figure*}
    \centering
    \includegraphics[width=\textwidth]{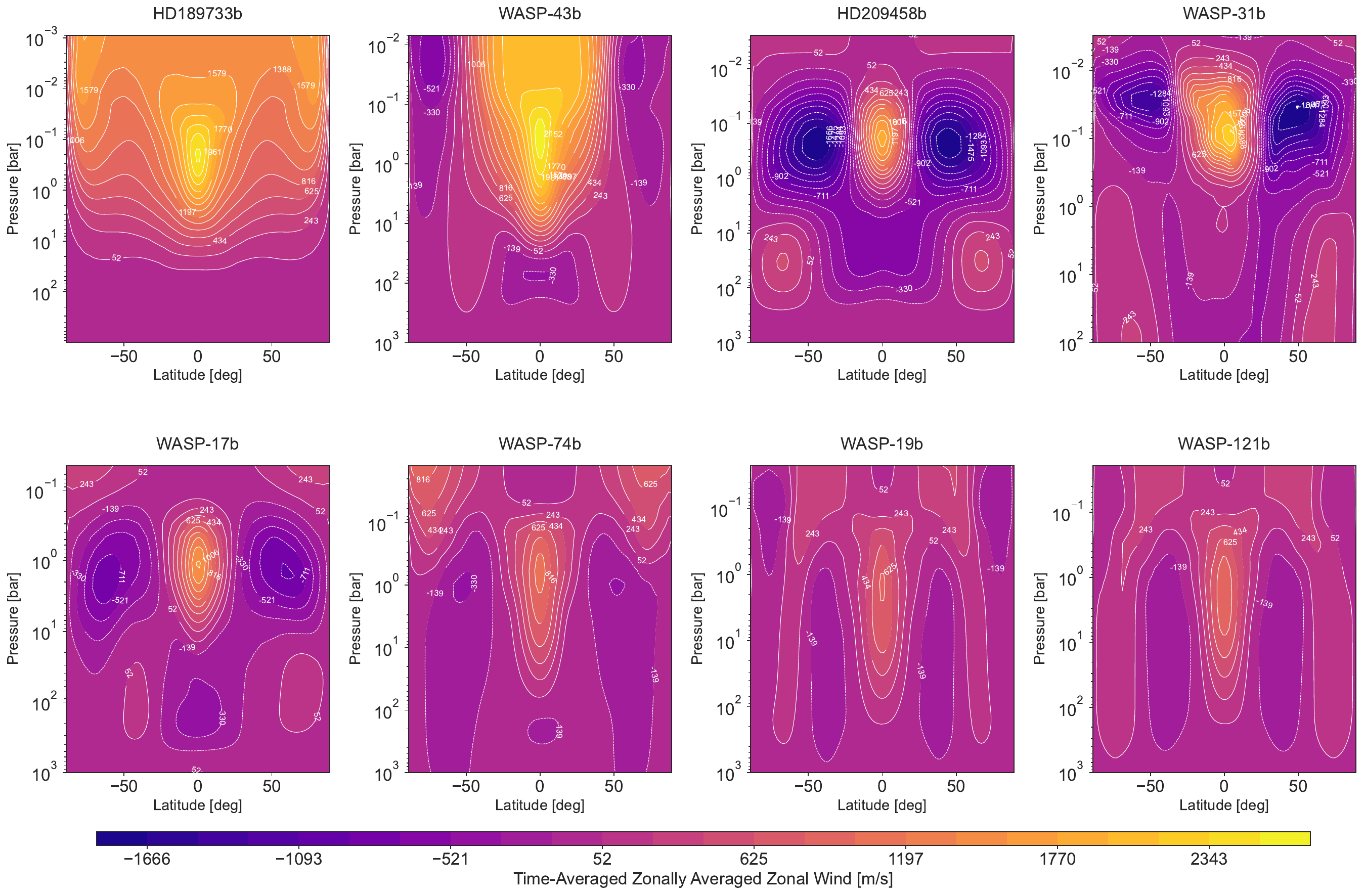}
    \caption{Zonally-averaged zonal winds for the 8 planet in our GCM sample. Plots are time-averaged over the last 500 days of their corresponding simulation times.}
    \label{fig:zonalmeanzonalwind}
\end{figure*}

Figure \ref{fig:zonalmeanzonalwind} shows the zonally-averaged zonal winds, revealing an equatorial jet feature in all simulations. As equilibrium temperature increases, the main jet feature at the equator weakens and split into multiple jets of alternating directions in higher latitudes.  In transitioning from HD 189733b (top-left panel) to hotter planets, the originally singular equatorial jet gradually shifts into deeper atmospheric layers. The upper layers, $P \leq 0.1~\mathrm{bar}$, particularly in the hotter cases shown on the bottom row, exhibit weakening winds. Notably, the peak wind speeds go from $\approx 2600~ \text{m}\cdot s^{-1}$ for HD 189733b to around $800~ \text{m} \cdot s^{-1}$ for WASP-121b. This phenomenon is further highlighted in Figure \ref{fig:los_terminatorslice_tavg_worotation}, which presents a single-layer snapshot of projected line-of-sight velocity viewed from the nightside. The observer's exact orientation is further detailed in Figure \ref{fig:visual_aid}. The same dynamics described in Figure \ref{fig:zonalmeanzonalwind} appear here: in HD 189733b, the entire evening terminator (spanning $P \approx 10^{-2}$ to $10~\mathrm{bar}$) exhibits negative line-of-sight velocities (i.e., flow toward the observer), while the morning terminator flows away from the observer. Peak velocities concentrate near the equator, consistent with the equatorial jet. As the equilibrium temperature increases from the top-left to the bottom-right panels, the main jet first narrows latitudinally (e.g., in WASP-43b) and later splits into multiple, alternating jets, with the equatorial jet becoming increasingly confined. In the transition from WASP-17b to WASP-121b (bottom row), the equatorial jet remains around $ P \approx 1~\mathrm{bar}$, whereas the upper layers ($P \leq 0.1~\mathrm{bar}$) develop a growing global sub-stellar-to-anti-stellar flow. As noted by \cite{showman_doppler_2013}, an internal return flow, visible in the deeper regions, is required to conserve mass and balance the stronger day-to-night circulation in the upper atmosphere. Although the interior circulation appears weaker, the higher mass flux in these denser layers compensates for that reduced flow.

\subsection{Global Flow Transition}
\label{sec:Global Flow Transition}
\begin{figure*}
    \centering
    \includegraphics[width=\textwidth]{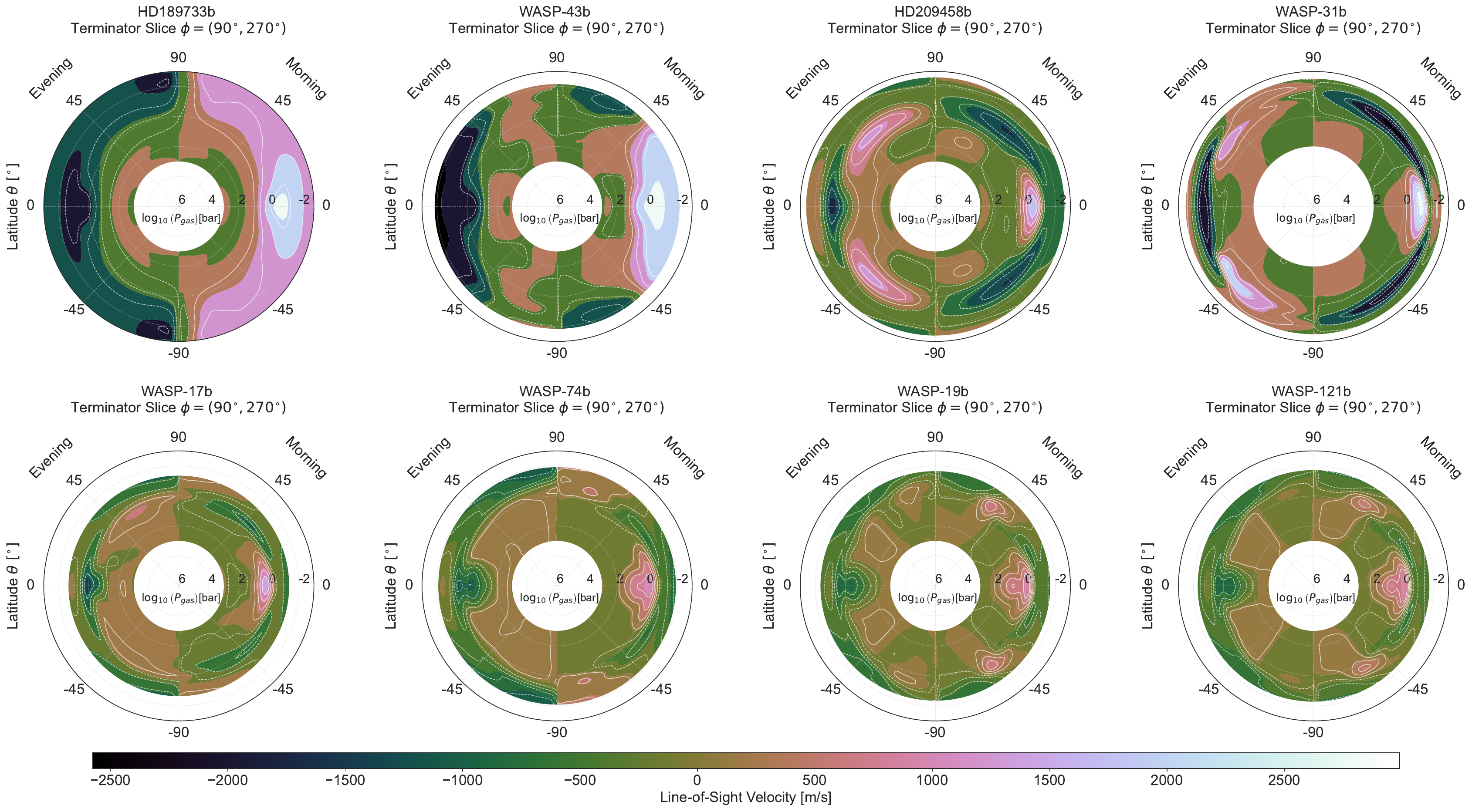}
    \caption{Line-of-sight velocities are presented across a slice along the terminator. Equilibrium temperature increases from the coldest hot Jupiter case in our sample HD~189733b on the upper left corner to the hottest hot Jupiter example in our sample WASP-121b on the bottom right corner. Positive values denote winds moving away from the observer and the effect of planetary rotation is not taken into account. For guidance in reading this plot please refer to Figure \ref{fig:visual_aid}.}
    \label{fig:los_terminatorslice_tavg_worotation}
\end{figure*}

In Figure \ref{fig:OLRulev0.1}, starting from WASP-17b and hotter planets, we observe an equatorial flow opposing the equatorial jet form at the $P \approx 0.1$ bar atmospheric layer that is plotted. This feature exists, to a limited extent, even in colder planets such as HD 209458b or WASP-31b although not as prominently as for the hotter planets. For colder planets, this transition presents itself as the shrinking of the latitudinal extent of the equatorial jet with westward flows forming in the high-latitude regions of their respective atmospheres. As presented in the previous section, the extent and specifics of this dynamical feature can be better observed in Figure \ref{fig:los_terminatorslice_tavg_worotation}. 

The $T$-$P$ profiles in Figure~\ref{fig:TP_tavg_init_final} highlight the gradual shift in global flow regimes as planetary equilibrium temperatures increase. We can interpret these changes by comparing our physical understanding of two contrasting scenarios: one dominated by an eastward equatorial jet, where heat deposited at the sub-stellar point travels eastward to the evening terminator and then around the night side to the morning terminator, and another characterized by day-to-night flow, which heats both terminators more symmetrically despite the underlying eastward mean flow due to the planet's rotation.

In HD~189733b, the morning terminator is even colder than the anti-stellar point, and the evening terminator follows the sub-stellar profile, consistent with an eastward superrotating jet distributing heat efficiently. For WASP-43b, the morning terminator and anti-stellar temperatures nearly coincide, while the evening terminator—though still above the global mean—is cooler, indicating a weakening of the equatorial jet. In the transitional HD~209458b, the morning terminator now exceeds the anti-stellar temperature, and the two terminators have nearly identical profiles. This implies the emergence of day-to-night winds that heat the morning limb, while the diminishing jet is evident from the cooler evening terminator, which lies below the global mean that is skewed towards higher temperatures by an increasingly hotter sub-stellar region. For the hottest planets (shown on the bottom row), the temperature difference between the sub-stellar and anti-stellar regions can exceed 2000\,K. Despite this intensified day-to-night gradient, terminator differences stabilize at a few hundred Kelvin, and the evening terminator gradually converges toward the global mean, in line with our expectations of a more symmetric heating for a day-to-night flow.

\begin{figure*}[ht]
    \centering
    \includegraphics[width=\textwidth]{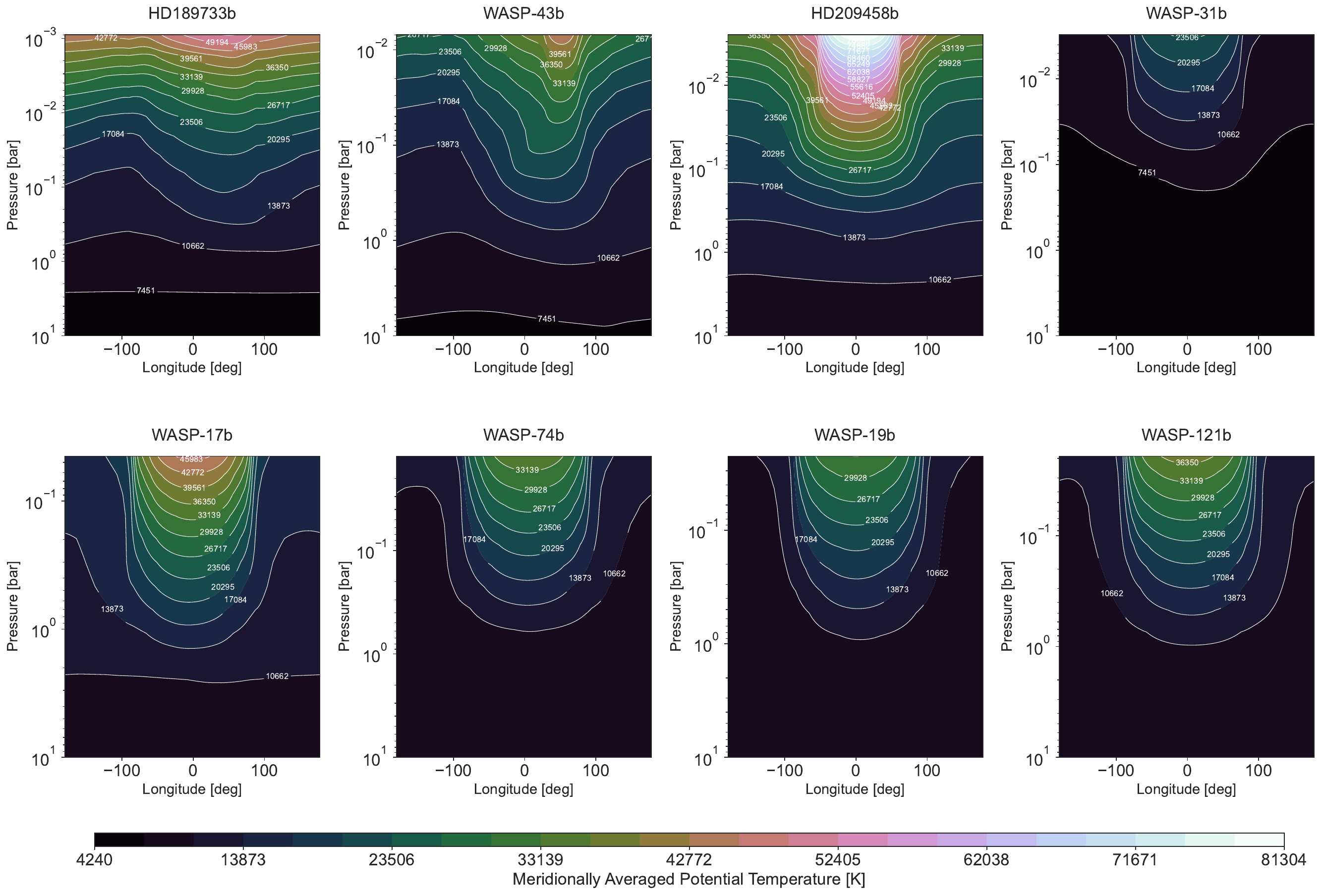}
    \caption{Latitudinally-averaged potential temperature plots. Sub-stellar point is located at $\phi=0$.}
    \label{fig:pt_mer_zoom}
\end{figure*}

As day-to-night flow patterns emerge in the hotter planets, we also see a strong connection between these circulation changes and the meridionally averaged potential temperature structure. While potential temperature is typically used to assess atmospheric stability and often plotted in a latitude-pressure framework \citep[e.g.][]{deitrick_thor2.0_2020}, here we examine it against longitude and pressure. This alternative representation not only confirms stable, vertically stratified profiles but also highlights longitudinal variations that correlate with the evolving circulation regimes. In Figure \ref{fig:pt_mer_zoom} we recover vertically stratified profiles as expected from convectively stable atmospheres. Beyond this baseline structure, two key findings stand out. First, HD 189733b and WASP-43b show a pronounced longitudinal asymmetry around the sub-stellar point at $0^{\circ}$, indicating significantly eastward shifted hotspots and strong zonal contrasts in potential temperature. Second, all planets feature longitudinal potential temperature gradients, hinting at baroclinic\footnote{We use the term “baroclinic” more generally: any misalignment between pressure surfaces and density surfaces represents a source of potential energy that can drive large-scale flow instabilities. On Earth, that misalignment primarily results from equator-to-pole temperature contrasts; on other planets—or in particular dynamical regimes—significant day-night gradients can play the same role.} instabilities. Since our vertical axis represents constant pressure levels, any horizontal gradient in potential temperature implies a corresponding density gradient, which can drive the instability. In contrast to the colder planets, the hotter planets exhibit more symmetric distributions down to $P \approx 0.5-1$ bar. This range corresponds to the vertical depth of their day-to-night flows, as illustrated in Figure~\ref{fig:los_terminatorslice_tavg_worotation}, which shows the extent of these winds in the upper layers. As equilibrium temperature increases, we observe the hotspot shift diminish and the symmetric region broaden, aligning with the intensified stellar irradiation and the growing dominance of day-to-night circulation.

\subsection{Benchmarking the Wavelet Analysis Method}
\label{sec:Benchmarking the Wavelet Analysis Method}
Having observed the emergence of a general flow transition pattern across our simulated hot Jupiter cases, we would like to understand the atmospheric waves that drive these patterns. For this purpose, we have introduced a novel wavelet-based analysis framework in Section \ref{section:Wavelet Analysis}. In order to decode the wave patterns underlying the large-scale flow transitions, we first benchmark our wavelet analysis method by decomposing time-dependent analytical solutions to the shallow water equations as they are presented in \citet{heng_analytical_2014}, but originally derived by \citet{matsuno_quasi-geostrophic_1966}. These analytical solutions enable us to verify that the method retrieves the correct wavenumbers and frequencies, while also exemplifying the kind of output our approach produces. As benchmarks, we use the solutions to the following dispersion relation (equation 106 in \citealt{heng_analytical_2014}),
\begin{equation}
    \omega_{R}^3 - \left(2n + 1 + k^2 \right) \omega_{R} - k =0
\end{equation}
where $\omega_{R}$ is the real part of the wave frequency, $n$ is the meridional wavenumber and $k$\footnote{In the current and following equations, adopted from \citet{heng_analytical_2014}, we denote the zonal wavenumber as $k$, which corresponds to $k_{x}$ in their work.} is the zonal wavenumber. This dispersion relation is derived under the equatorial $\beta$-plane approximation and the assumption of a balanced, frictionless flow ($\omega_{I}=0$). The general time-dependent solutions (equation 107 in their work) provide an exact, analytical benchmark for our wavelet analysis method. We adopt the algebraic notation of \citet{heng_analytical_2014}, while again noting that these solutions were originally derived by \citet{matsuno_quasi-geostrophic_1966} and expanded upon by \citet{ gill_simple_1980}:

\begin{align}
v'_ {y} &= v_{0} ~\exp \left(-  \frac {y^ {2}}{2} \right)  ~\mathcal{H}_{n} \cos \left( k x - \omega_{R} t \right) ,
\end{align}

\begin{multline}
h' = v_{0}  ~\exp \left( - \frac {y^{2}}{2} \right) \left( \frac{k}{\omega_{R}} - \frac{\omega_{R}}{k} \right)^{-1} \sin \left( k x - \omega_{R} t \right) \times \\
\left[y \left( \frac{1}{k} + \frac{1}{\omega_{R}}\right) ~\mathcal{H}_{n} - \frac{2n ~\mathcal{H}_{n-1} }{k}\right],
\end{multline}

\begin{multline}
v'_{x} = \frac{v_{0}}{k^{2} - \omega_{R}^{2}} ~\exp \left( - \frac{y^{2}}{2} \right) \sin \left( k x - \omega_{R} t \right) \times \\
\left[ y \left( \omega_{R} + k \right) ~\mathcal{H}_{n} - 2n k ~\mathcal{H}_{n-1}\right],
\end{multline}
where $t$ is time, $v_{0}$ is an arbitrary normalization constant, $v'_{x}$ are zonal velocity perturbations, $h'$ shallow water height perturbations, $v'_ {y}$ meridional velocity perturbations, and $\mathcal{H}_{n}$ Hermite polynomials of order $n$. In the equatorial $\beta$-plane approximation, $x$ and $y$ are the horizontal and vertical distances in Cartesian coordinates, which correspond to latitude and longitude on a sphere.

\begin{figure*}
    \centering
    \includegraphics[width=\textwidth]{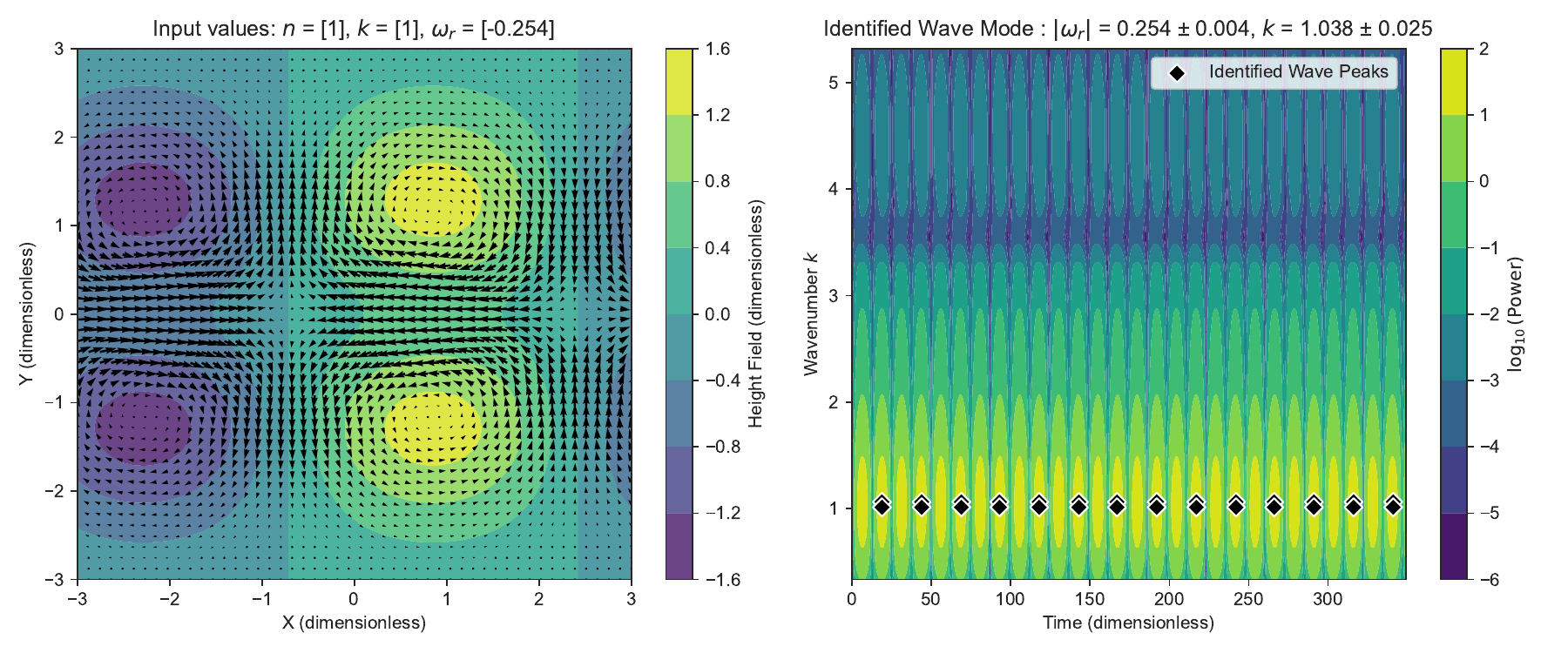}
    \caption{A single analytical wave benchmark example for our wavelet analysis method. \textbf{Left}: A Matsuno-Gill type analytical solution that corresponds to a Rossby wave characterized by $n = 1 ~;~ k = 1$. Plotted are the zonal wind components overlaid on the shallow water height field. \textbf{Right}: A scalogram of the zonal wavenumber against dimensionless time. Our method correctly identifies the input zonal wavenumber $k$ and the temporal frequency $\omega_{R}$ consistent within the uncertainties, which is due to the Heisenberg-Gabor uncertainty principle.}
    \label{fig:analytical_identification}
\end{figure*}

Figure \ref{fig:analytical_identification} displays a single analytical Rossby wave\footnote{For clarification, the Rossby waves referenced in the main text refer to Rossby-type solutions to the shallow water equations under the equatorial $\beta$-plane approximation, as described in \citet{matsuno_quasi-geostrophic_1966}.} solution to the shallow water equations on the left panel. The right panel of the same figure displays the time-dependent power spectrum of the zonal wind field at $y = 0$, corresponding to the equator, calculated utilizing the wavelet analysis method. To calculate the zonal wavenumber, we apply the wavelet decomposition to the zonal velocity perturbations $v'_{x}$ across $x$ at each $y$ and then repeat it for all $y$-values. This decomposition identifies a band in wavenumber space with oscillating power. The spread of the wavenumber band displays the previously mentioned (see Section \ref{section:Wavelet Analysis}) inherent Heisenberg-Gabor uncertainty, which limits the simultaneous precision of the recovered wavenumbers and their spatial localization. Analyzing the periodicity of the identified wavenumber $k$ allows us to extract the temporal frequency $\omega_{R}$. Our recovered values for the presented example $k = 1.038 \pm 0.025  ~\text{and} ~ |\omega_R| = 0.254 \pm 0.004 $, as seen in Figure \ref{fig:analytical_identification}, are consistent with the input values $n = 1 ~\text{and}~ k = 1$. Our analysis recovers the absolute value of $|\omega_R| = 0.254$, but not its sign as the wavelet method is unable to extract the travel direction of the wave. While we illustrate results at a single $y$-location, repeating the same analysis for different $y$-values or averaging the outcomes over $y$ yields consistent results (not shown). This is an expected outcome given the simplicity of this single-wave example.

\begin{figure*}
    \centering
    \includegraphics[width=\textwidth]{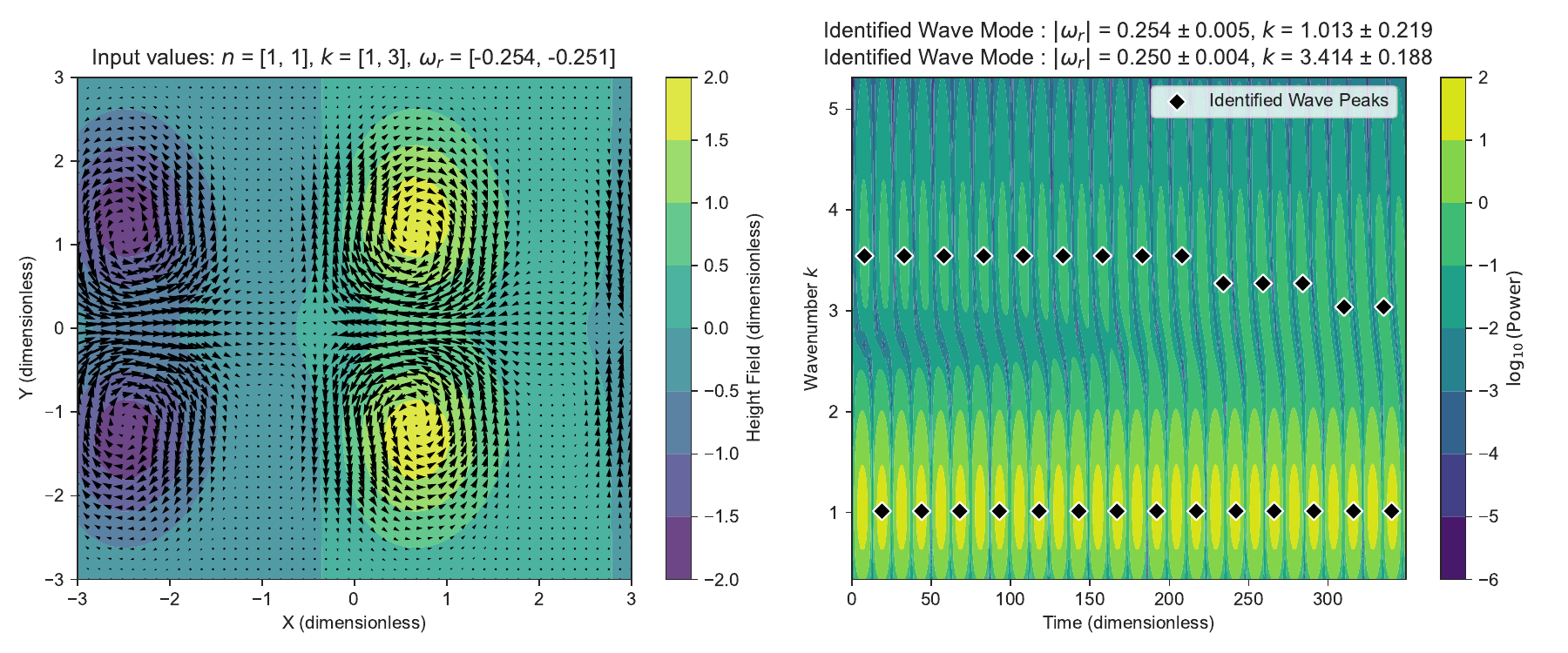}
    \caption{A multi-wave analytical benchmark example for our wavelet analysis method. \textbf{Left}: A Matsuno-Gill type analytical solution that corresponds to a combination of Rossby waves characterized by $n = 1, 1 ~;~ k = 1, 3 $. Plotted are the wind fields overlaid on the shallow water height field. \textbf{Right}: A scalogram of the zonal wavenumber against dimensionless time. Our method correctly identifies the input zonal wavenumbers $k$ and the temporal frequencies $\omega_r$ within the uncertainties, which is due to the Heisenberg-Gabor uncertainty principle.}
    \label{fig:analytical_identification_multi}
\end{figure*}

When multiple waves with different $k$ values are present, they can partially couple with one another, exacerbating the spread in the multiple identified wavenumbers. As an example, illustrated on the left panel of Figure~\ref{fig:analytical_identification_multi}, we introduce two Rossby wave solutions with $n = \left[1, 1\right] ~\text{and}~ k = \left[1, 3\right] $, which correspond to temporal frequencies $\omega_R = - \left[0.254, 0.251\right]$. As shown on the right panel of  Figure~\ref{fig:analytical_identification_multi}, our wavelet analysis recovers two distinct wave modes with $k = \left[1.013 \pm 0.219,\; 3.414 \pm 0.188\right]
\quad \text{and} \quad |\omega_R| = \left[0.254 \pm 0.005,\; 0.250 \pm 0.004\right]$. These values align closely with the input, despite the expected inherent spread introduced by Heisenberg-Gabor uncertainty and the wave-mode coupling. Overall, these benchmark tests demonstrate that our wavelet method is capable of accurately extracting wave modes from a flow field.

\subsection{Applying Wavelet Analysis to GCM Outputs}
\label{section: Applying Wavelet Analysis to GCM outputs}
In the previous section, we demonstrated our wavelet analysis method by recovering known wave structures from linear shallow-water solutions. This benchmarking served two key purposes: it verified our ability to identify imposed modes (e.g., specific Rossby waves) and introduced the outputs of a wavelet transform, particularly the scalograms\footnote{Scalograms are time-dependent power spectra produced by a wavelet transform. For an introduction to wavelet transform terminology and applications, see \citet{ten_lectures_on_wavelets_1993}.}, which show the time-evolving power in different modes. Having established the validity of our approach, we now turn to the more complex, non-linear flow solutions produced by our GCM.

Unlike the shallow-water test cases, GCM solutions can feature wave interactions that vary significantly with latitude. As also explained in \citet{mendonca_2020}, we expect Kelvin waves to dominate near the equator, where they appear as eastward-traveling features with vanishing meridional components. By contrast, Rossby waves generally vanish near the equator (where the Coriolis force is zero) but emerge at mid- and high latitudes, exhibiting anti-symmetric meridional structures. Both wave types are expected to show symmetry about the equator in temperature and zonal wind fields. Consequently, we analyze the wavelet signals separately at the equator and mid-latitudes to capture each wave’s contributions. Beyond the scalograms, we also examine the spatial distributions of the wavelet coefficients. By mapping power in a single wavenumber across latitude and longitude at a given atmospheric layer, we can confirm if a mode’s structure and symmetry match the expected signatures of Kelvin–Rossby waves. This combined perspective lets us connect each identified mode to a physically interpretable wave structure.

\begin{figure*}
    \centering
    \includegraphics[width=\textwidth]{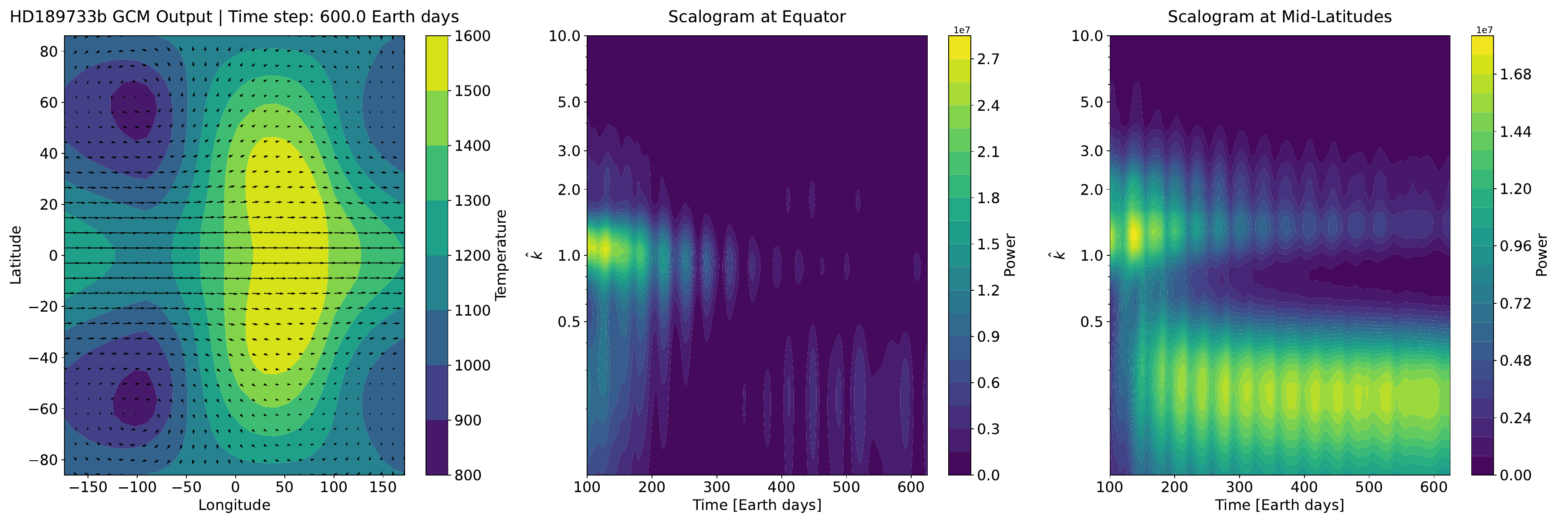}
    \caption{\textbf{Left panel:} Snapshot of the GCM solution for HD~189733b at 600 days, showing temperature and wind fields at approximately 0.1~bar.  \textbf{Center \& Right:} Wavelet scalograms of the wind field at the equator and mid-latitudes, showing prominent $\hat{k}=1$, $\hat{k}=2$ and developing $\hat{k} = 0.3$ modes.}
    \label{fig:hd189b_wavelet_decomp}
\end{figure*}

\begin{figure*}
    \centering
    \includegraphics[width=\textwidth]{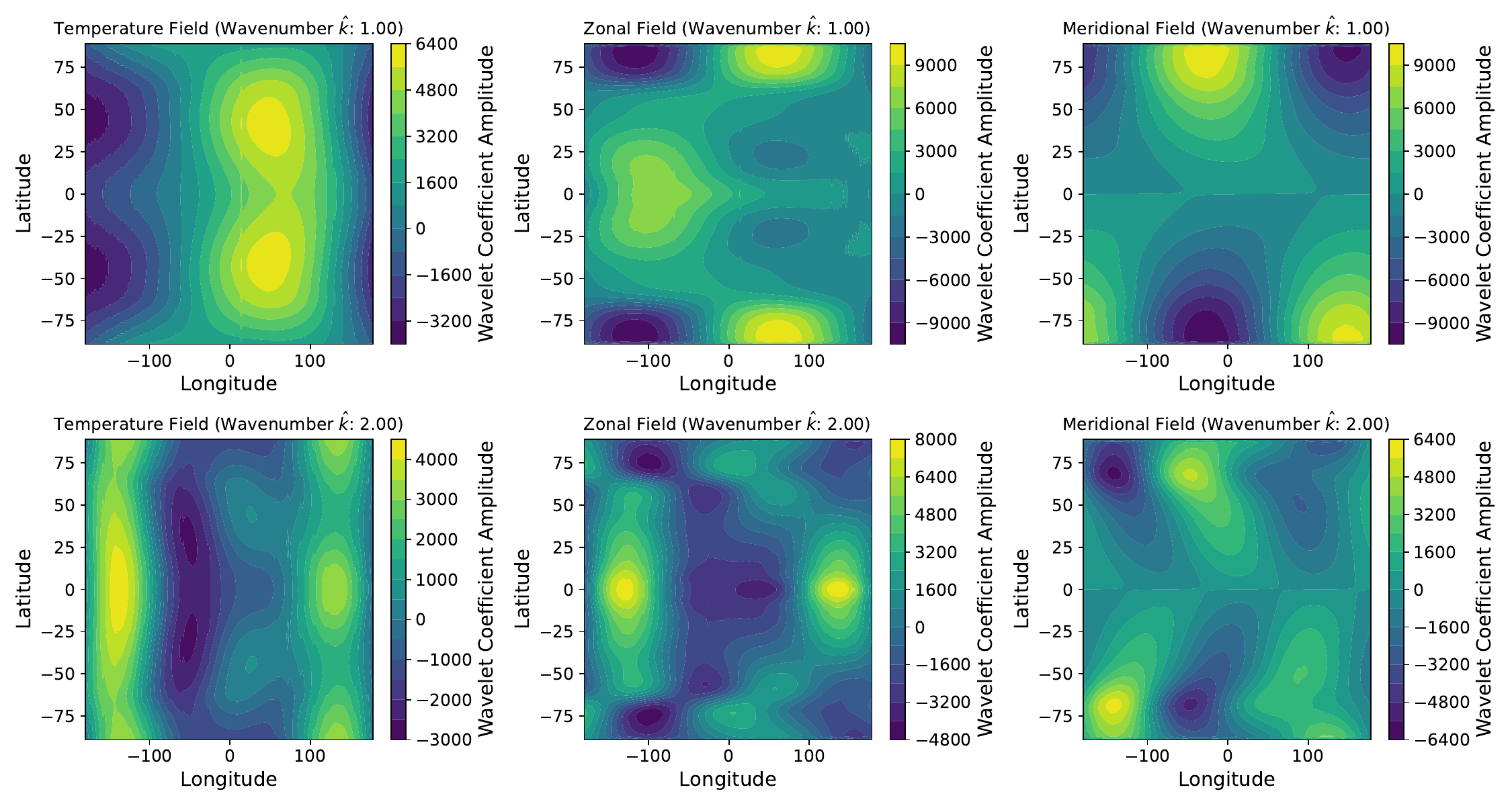}
    \caption{Spatial distribution of wavelet coefficients for the temperature, zonal wind and meridional wind fields of HD~189733b across the $\approx 0.1$ bar atmospheric layer. The shown results are time-averaged over the last 100 days of the analysis period. \textbf{Top} and \textbf{bottom} panels show the power in modes $\hat{k}=1$ and $\hat{k}=2$, respectively.}
    \label{fig:hd189b_k_1_2}
\end{figure*}

Figures~\ref{fig:hd189b_wavelet_decomp} and~\ref{fig:hd189b_k_1_2} illustrate this analysis for HD~189733b. In Figure~\ref{fig:hd189b_wavelet_decomp}, the left panel shows a snapshot of the GCM output at the $P \approx 0.1$\,bar layer (corresponding to the IR photosphere) at 600~days. The middle and right panels display the scalograms at the equator and mid-latitudes, respectively, following the approach used in Section \ref{sec:Benchmarking the Wavelet Analysis Method}. Figure~\ref{fig:hd189b_k_1_2} presents the spatial distribution of wavelet coefficients of the temperature, zonal wind, and meridional wind fields (from left to right) at the same pressure layer for wavenumbers $\hat{k}=1$ and $\hat{k}=2$, averaged over the final 100 days of the 500-day period, thereby providing a link between identified wave modes and physically anticipated wave structures. 

Wavenumbers $\hat{k} \approx 0.3$, $\hat{k} \approx 1$, and $\hat{k}=2$ consistently appear in our analyses and form an instructive basis for discussion. By definition (Section~\ref{section:Wavelet Analysis}), $\hat{k}=1$ corresponds to a planetary-scale wave; hence, we classify $\hat{k} \approx 0.7\text{–}1.5$ as planetary-scale, $\hat{k} \geq 1.5$ as small-scale, and $\hat{k} \leq 0.7$ as large-scale. We decompose and analyze the simulation output from 100 to 600 days after the start of each run, covering a part of the spin-up phase. The output frequency is set to 0.1~days—less than $25\%$ of the rotation period for all hot Jupiters in our sample (0.81–3.74~days)—to, assuming a Nyquist-like sampling, adequately resolve the oscillations in temperature, zonal wind, and meridional wind. Although applying the same wavelet decomposition at a later time, closer to steady-state, would still identify the waves, omitting the spin-up period would overlook essential details of how they form and interact with the evolving background flow—information that is important for understanding the wave-mean flow interplay that ultimately shapes the global circulation. 

For HD~189733b, we observe a well-developed zonal jet in the left panel of Figure~\ref{fig:hd189b_wavelet_decomp}. The middle panel (the equatorial scalogram) shows significant power in the planetary-scale $\hat{k} = 1$ mode, which diminishes over time and gives way to developing large-scale structures ($\hat{k} \approx 0.3$) around the 350-day mark. A weak contribution from the small-scale $\hat{k}=2$ mode is also visible. The right panel (the mid-latitude scalogram) shows a similar pattern: the power initially concentrates in wavenumbers $\hat{k} = 1 \text{-} 2$ and, although it fades over time as in the equatorial case, it remains present throughout the analysis period. Meanwhile, a large-scale flow with $\hat{k} \approx 0.3$ emerges. We interpret these large-scale modes as the mean flow modulated by wave–mean flow interactions. The overlap in power near the 170-day mark, coupled with a gradual decrease in planetary- and small-scale power and a corresponding increase in large-scale power, suggests a coupling with the $\hat{k}=0.3$ and $\hat{k}=1\text{-}2$ modes. This coupling aligns with the onset of equatorial superrotation, seen in the gradual increase of the large-scale mode in the equatorial scalogram.

Inspecting Figure~\ref{fig:hd189b_k_1_2} reinforces our interpretation since both wave modes $\hat{k}=1$ and $\hat{k}=2$  display characteristics of coupled Rossby–Kelvin waves: symmetric temperature and zonal-wind structures about the equator, coupled with anti-symmetric meridional features at higher latitudes that vanish at the equator. Additionally, the upper middle panel features an equatorial feature reminiscent of the Matsuno-Gill \citep{matsuno_quasi-geostrophic_1966, gill_1982} chevron pattern. Altogether, these observations paint a picture that is consistent with the \citet{showman_polvani_2011} explanation of how equatorial superrotation arises. Namely, we expect momentum to be transported from higher latitudes onto the equator through the coupling of westward-propagating Rossby waves and eastward-propagating equatorial Kelvin waves, creating the characteristic chevron shape.

\begin{figure*}
    \centering
    \includegraphics[width=\textwidth]{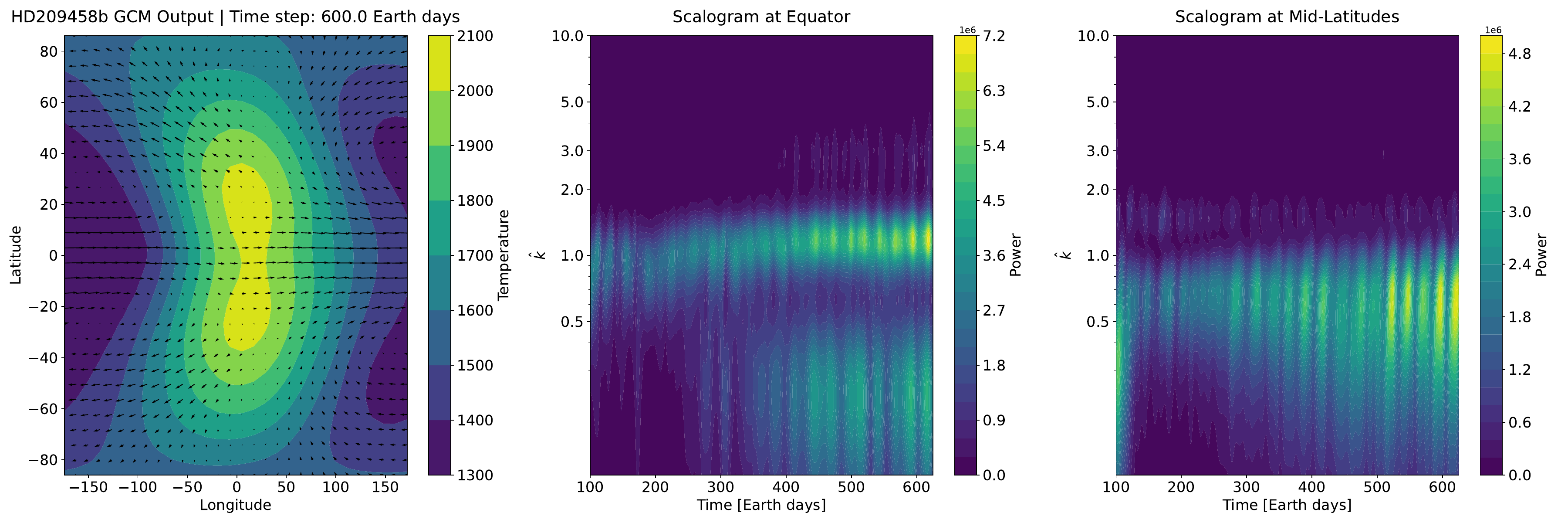}
    \caption{\textbf{Left panel:} Snapshot of the GCM solution for HD~209458b at 600 days, showing temperature and wind fields at approximately 0.1~bar. \textbf{Center \& Right:} Wavelet scalograms of the wind field, illustrating a qualitatively similar case to HD~189733b, albeit with a weakening $\hat{k}=2$ contribution.}
    \label{fig:hd209b_wavelet_decomp}
\end{figure*}

HD~209458b (Figure~\ref{fig:hd209b_wavelet_decomp}) exhibits a transitional flow regime. On the left panel, we observe a narrower zonal jet compared to HD 189733b, and mid-to-high-latitude westward flows become more pronounced. The middle panel displays a significant $\hat{k} \approx 1$ mode that persists throughout the analysis period, with slightly increasing power. Meanwhile, a large-scale ($\hat{k} \approx 0.3$) structure develops around the 300-day mark, accompanied by a later weak contribution from smaller scales ($\hat{k} \approx 3$). A similar pattern appears in the right panel, where features of $\hat{k} \approx 1$ and $\hat{k} \approx 2$ persist over the entire analysis period. The $\hat{k}\approx 1$ feature broadens to span $\hat{k} \approx 0.3\text{–}1$, peaking near $\hat{k} \approx 0.5\text{–}0.7$. The development of this broadening coincides with the emergence of the large-scale feature in the equatorial scalogram, suggesting a coupling of the identified waves similar to what is observed in HD 189733b.

\begin{figure*}
    \centering
    \includegraphics[width=\textwidth]{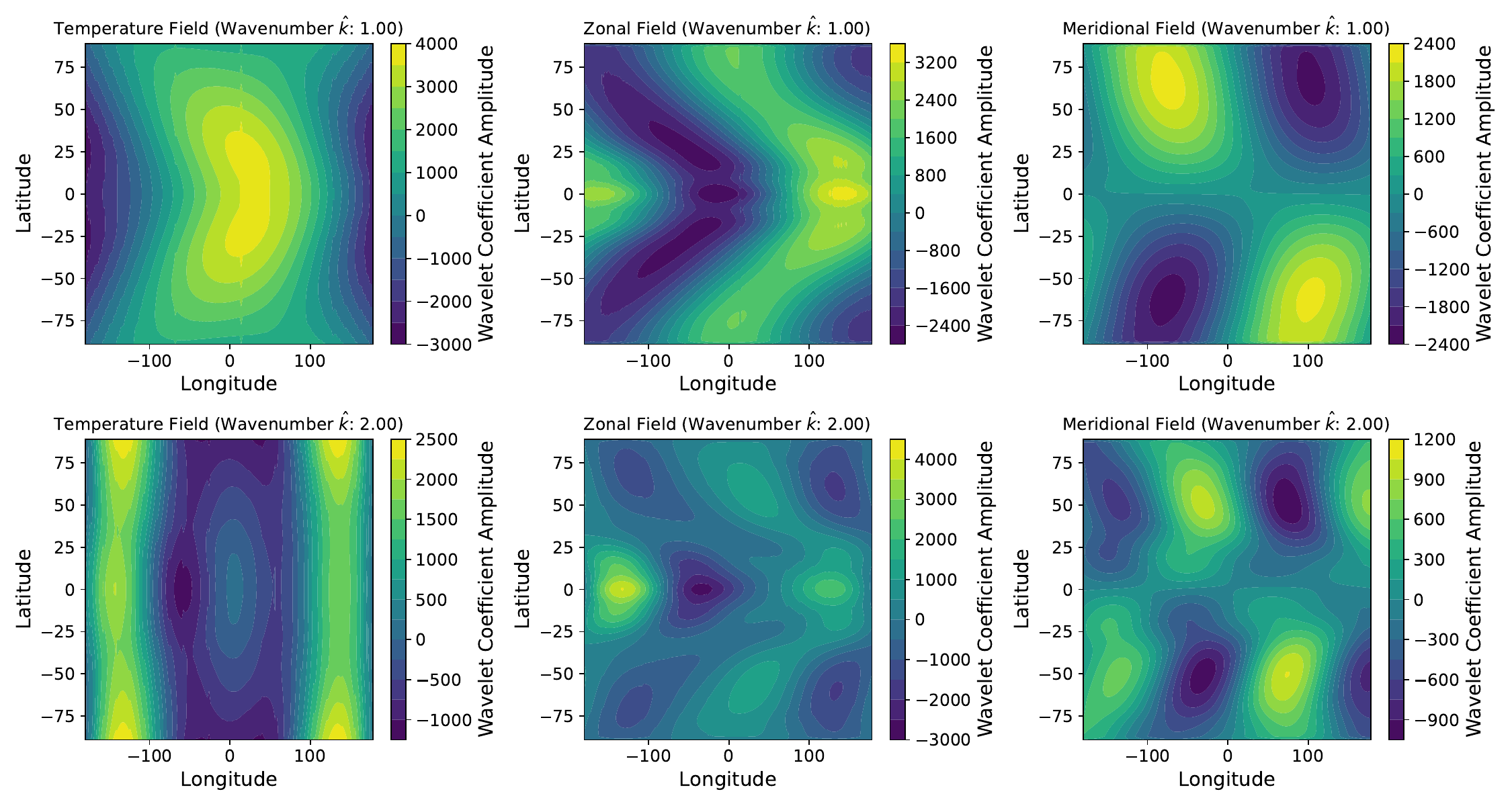}
    \caption{Spatial distribution of wavelet coefficients for the temperature, zonal wind and meridional wind fields of HD~209458b across the $\approx 0.1$ bar atmospheric layer. The shown results are time-averaged over the last 100 days of the analysis period. \textbf{Top} and \textbf{bottom} panels show the power in modes $\hat{k}=1$ and $\hat{k}=2$, respectively.}
    \label{fig:hd209b_k_1_2}
\end{figure*}

In Figure~\ref{fig:hd209b_k_1_2}, the $\hat{k}=1$ and $\hat{k}=2$ modes retain Rossby–Kelvin wave characteristics: temperature and zonal-wind structures remain symmetric about the equator, while meridional winds are anti-symmetric. The meridional wind components still vanish at the equator, displaying the expected Kelvin-wave feature, but this vanishing is confined to within $\pm 10^\circ$ of the equator, compared with $\pm 30^\circ$ for HD 189733b in Figure \ref{fig:hd189b_k_1_2}. In addition, substantial phase tilts appear at higher latitudes for both the $\hat{k}=1$ and $\hat{k}=2$ modes. The tilted zonal wind features, forming an eastward-pointing chevron shape near the equator, result from the coupling between eastward-propagating Kelvin waves and westward-propagating Rossby waves. However, in HD 209458b, the effect appears weaker than in HD 189733b.

\begin{figure*}
    \centering
    \includegraphics[width=\textwidth]{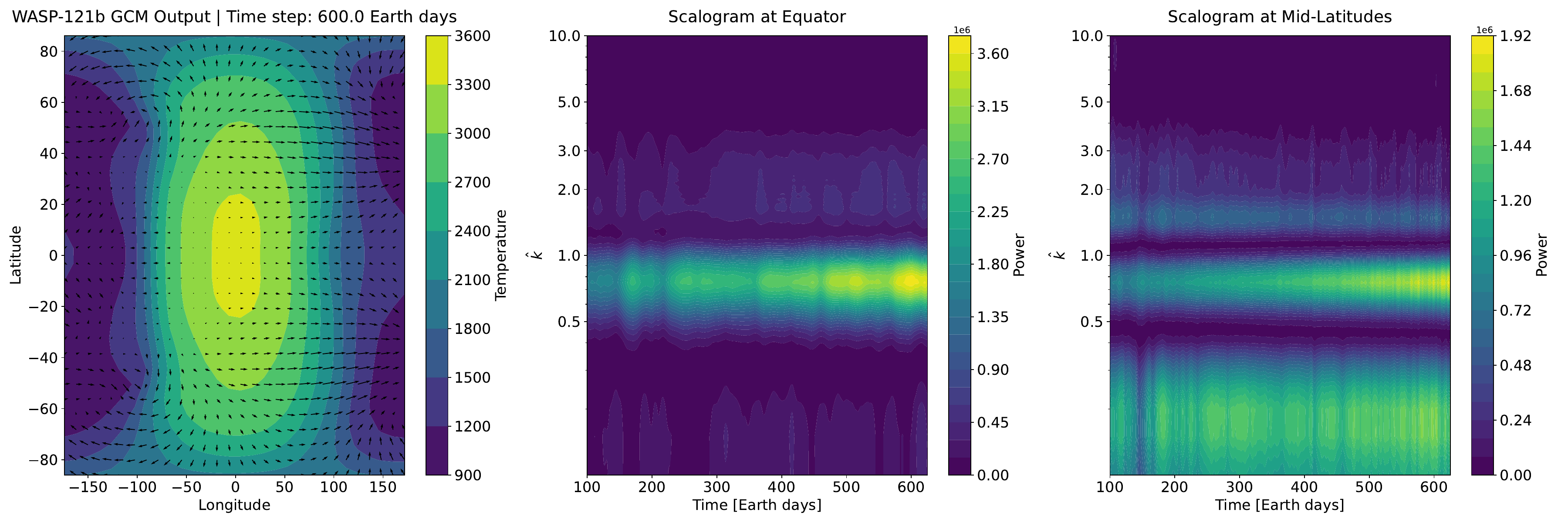}
    \caption{\textbf{Left panel:} Snapshot of the GCM solution for WASP-121b at 600 days, showing temperature and wind fields at approximately 0.1~bar. \textbf{Center \& Right:} Scalograms of the zonal wind at the equator and mid-latitudes reveal a prominent $\hat{k} \approx 1$ mode in both regions. Large-scale ($\hat{k} \leq 0.5$) and small-scale ($\hat{k} \geq 1.5$) contributions are also evident in both, though they are weaker at the equator.}
    \label{fig:wasp121b_wavelet_decomp}
\end{figure*}

Finally, WASP-121b (Figure~\ref{fig:wasp121b_wavelet_decomp}) displays a fully developed day-to-night flow in the left panel. Both the equatorial and mid-latitude scalograms share qualitatively similar patterns, with the strongest signal given by the $\hat{k} \approx 1$ mode. Alongside the planetary-scale wave modes, $\hat{k} \approx 0.3$ and $\hat{k} \approx 2$ are present at both the equator and mid-latitudes, and both modes appear stronger at mid-latitudes. The contributions from each wave mode remain steady during the analysis, showing no signs of evolution. Since, at this stage of the simulation, we do not expect the solutions to have reached a steady-state, we interpret these modes as the wave response to the instellation and rotation of the planet that do not display any signs of interactions.

\begin{figure*}
    \centering
    \includegraphics[width=\textwidth]{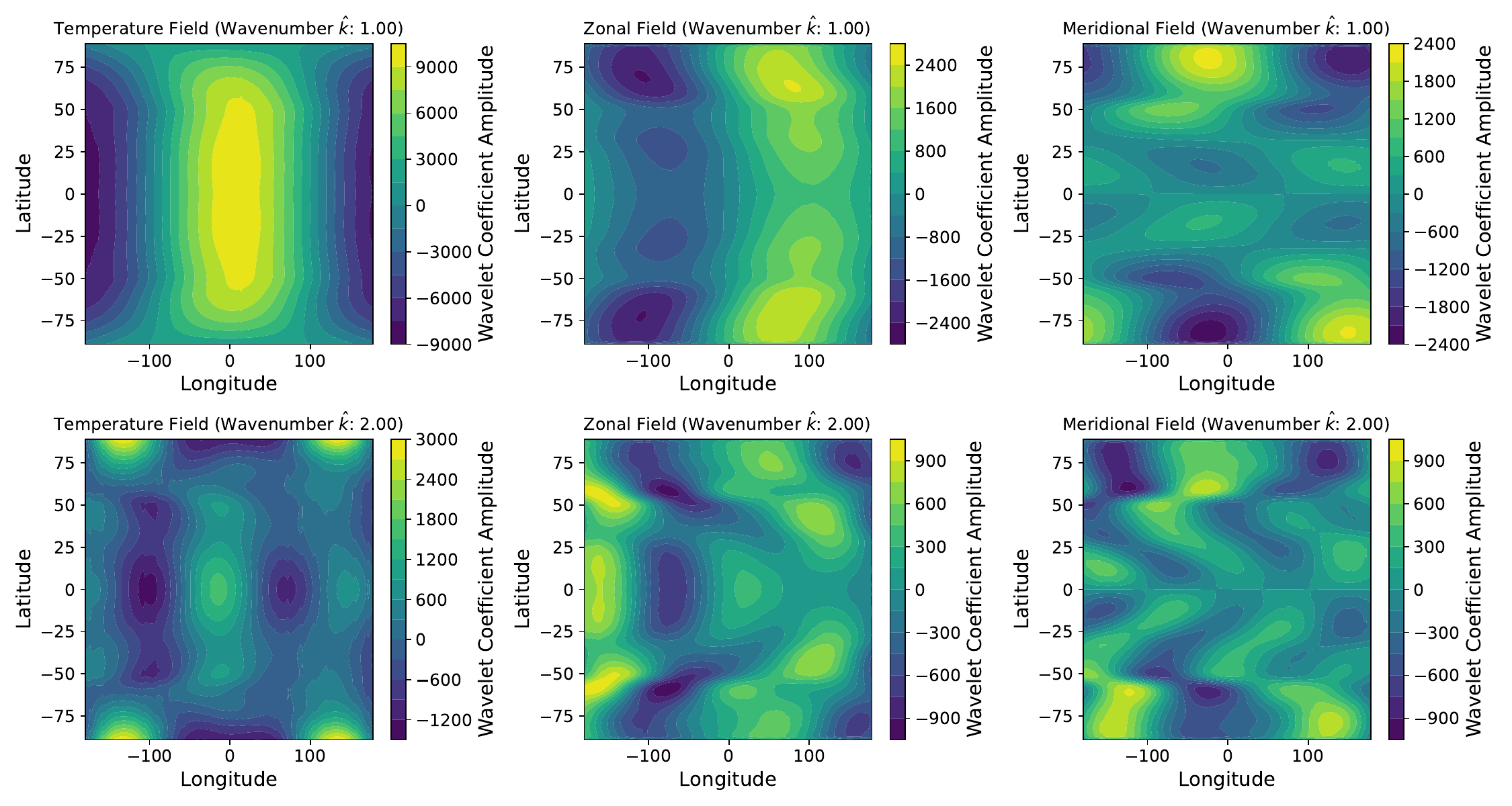}
    \caption{Spatial distribution of wavelet coefficients for the temperature, zonal wind and meridional wind fields of WASP-121b across the $\approx 0.1$ bar atmospheric layer. The shown results are time-averaged over the last 100 days of the analysis period. \textbf{Top} and \textbf{bottom} panels show the power in modes $\hat{k}=1$ and $\hat{k}=2$, respectively.}
    \label{fig:wasp121b_k_1_2}
\end{figure*}

Figure~\ref{fig:wasp121b_k_1_2} shows that the $\hat{k}=1$ mode lacks strong equatorial signatures. Contributions appear at higher latitudes, where weak Rossby-wave signals persist without coupling to an equatorial Kelvin wave. This pattern aligns with circulation dominated by direct stellar forcing rather than wave–mean flow interactions. By contrast, the $\hat{k}=2$ panels exhibit some equatorial features, with meridional wind components vanishing only at the equator. This implies that the wave response still gets triggered at smaller scales, but is unable to grow to form planetary-scale dynamical features. The same high-latitude ($\pm60^\circ$) phase tilts observed on HD~209458b also occur here. Moreover, the spatial distributions divide into two regions above and below $\pm60^\circ$, coinciding with the temperature feature seen in the upper-left panel, which corresponds to the hotspot. We interpret this boundary as marking the transition from a region with very short radiative timescales that disrupts dynamical formation to the high-latitude zone, where lower temperatures still allow dynamical structures to form.

\section{Discussion}
\label{section:Discussion}
\subsection{Importance of coupling small- and large-scale modes}
The evolution of the dominant wavenumber modes found in Section \ref{section: Applying Wavelet Analysis to GCM outputs} aligns with the theoretical expectations of \citet{showman_polvani_2011}, \citet{showman_doppler_2013}, \citet{showman_tan_parmentier_2020} and the modeling efforts of \citet{mayne_2017}. These studies suggest that equatorial superrotation arises from the coupling of smaller-scale eddies to the mean flow, with $\hat{k} \approx 1$ serving as the primary mode for such atmospheres. 

As proposed by \citet{mendonca_2020}, however, the $\hat{k} = 2$ mode also plays a crucial role, representing the smaller-scale eddy features that help drive the equatorial superrotation. Although a one-to-one mapping between wavenumber and physical process is unfeasible, focusing on the spin-up phase lets us analyze the atmosphere’s initial response to thermal forcing and planetary rotation, from which we can draw several conclusions. In all of our shown examples in Section \ref{section: Applying Wavelet Analysis to GCM outputs} (Figures \ref{fig:hd189b_wavelet_decomp}, \ref{fig:hd209b_wavelet_decomp}, \ref{fig:wasp121b_wavelet_decomp}), the $\hat{k}=1$ mode consistently appears strongest first as the simulation spins up, followed by the later emergence of the $\hat{k}=2$ and $\hat{k}=0.3$ modes. We propose that the initial emergence of the $\hat{k}=1$ mode reflects a planetary-scale response to thermal forcing. This is supported by the equatorial scalograms (Figures \ref{fig:hd189b_wavelet_decomp}, \ref{fig:hd209b_wavelet_decomp}, \ref{fig:wasp121b_wavelet_decomp}), which show slightly stronger $\hat{k}=1$ coefficients than those at mid-latitudes, consistent with the equatorial region being hotter and thus exhibiting a stronger response to thermal forcing, and by the steady-state analysis for WASP-121b (Figure \ref{fig:wavelet_last_500}), where $\hat{k}=1$ is the only remaining wave mode in the mid-latitudes due to intense instellation suppressing dynamic features. Looking at the same figure, we note that for HD 189733b and HD 209458b, where a zonal jet forms, the  $\hat{k}=1$ mode is absent or strongly weakened. This observation aligns with our physical expectations since efficient global heat redistribution would inhibit further excitation of the planetary-scale mode. The large-scale mode ($\hat{k} \approx 0.3$) appears to trace both the initial eastward background flow from the planetary rotation and the later developing equatorial superrotation. Additionaly, the initial time-dependent behavior offers an insight into how these dynamical features interact with each other. The sequence of events suggests that initial instellation-driven thermal forcing sets the stage for wave formation, and only then do the smaller-scale features develop and couple to the main flow. 

Our results also show that as equilibrium temperature increases, the contribution of the $\hat{k}=2$ mode weakens. Although this mode remains an important indicator of Rossby and equatorial Kelvin wave activity (i.e. equatorial features with vanishing meridional winds and distinct high‐latitude phase tilts), in hotter planets (e.g. WASP-121b) it does not couple effectively with the main flow to drive equatorial superrotation. In other words, while the $\hat{k}=2$ mode in cooler atmospheres can help transport momentum equatorward to drive the formation of the equatorial jet, in hotter planets the incomplete coupling of these small-scale eddies results in the emergence of a dominant day‐to‐night circulation pattern. This argument is further supported by the fact that the steady-state analysis of HD 189733b (Figure \ref{fig:wavelet_last_500}), which shows a persistent $\hat{k}=2$ contribution suggesting that this mode is responsible for sustaining the existing jet.

We further interpret this shift as reduced circulation efficiency under stronger instellation. When the radiative timescale is shorter than the wave response, small-scale eddies cannot couple effectively with the main flow to sustain dynamical structures. This is evidenced by the zonal phase tilts in WASP-121b (Figure \ref{fig:wasp121b_k_1_2}), which reveal Rossby-wave features that are not coupled to equatorial Kelvin waves (see Section \ref{section: Applying Wavelet Analysis to GCM outputs}). Following the reasoning by \citet{showman_polvani_2011}, such incomplete coupling limits the equatorward transport of angular momentum, resulting in a persistent day‐to‐night wind pattern that is mainly driven by the thermal forcing of the atmosphere.

\subsection{An ultra-hot Jupiter case: Modified WASP-121b runs}
\begin{figure*}
    \centering
    \includegraphics[width=\textwidth]{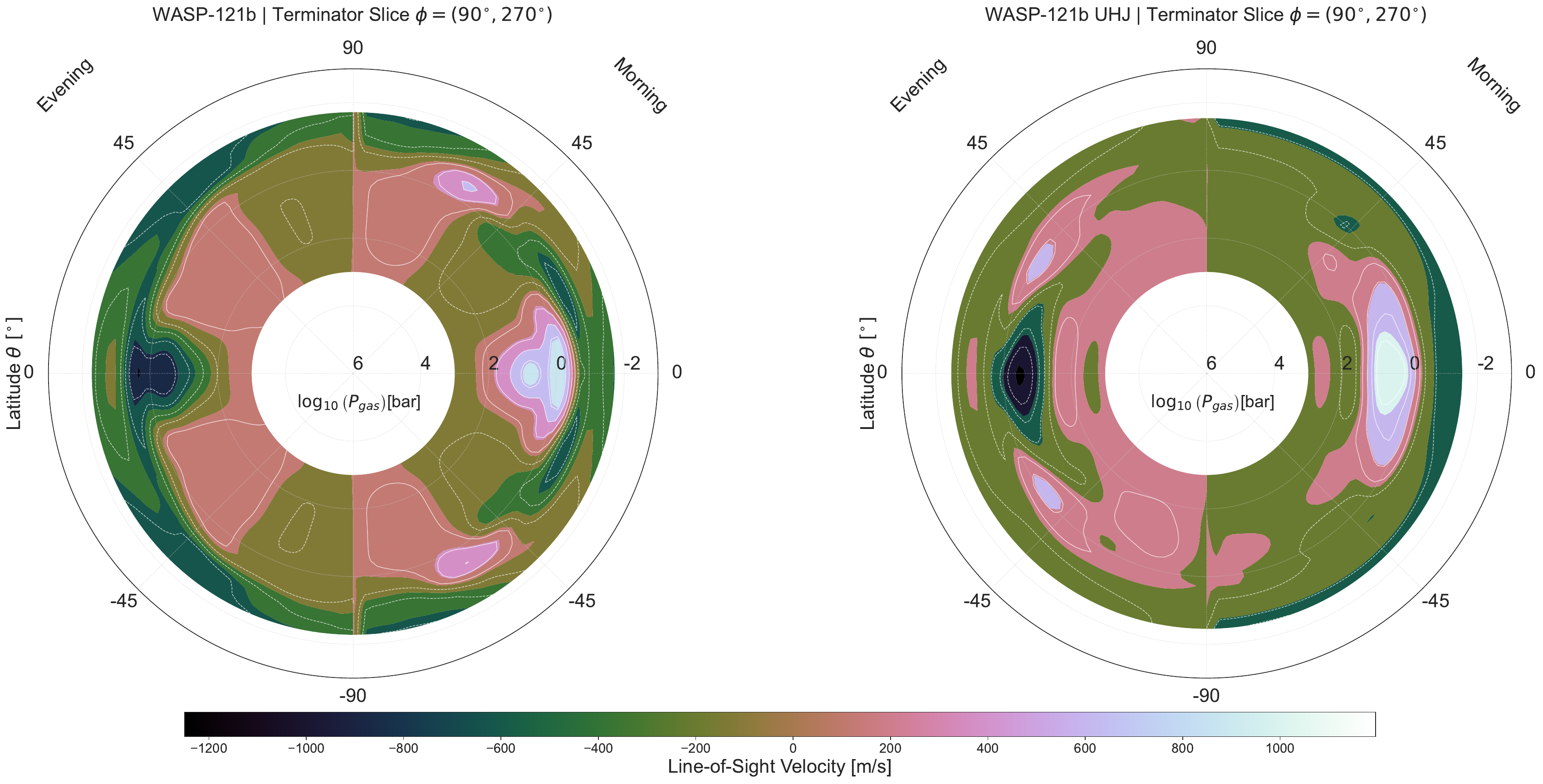}
    \caption{Line-of-sight velocities are presented across a slice along the terminator. Positive values denote winds moving away from the observer and the effect of planetary rotation is not taken into account. For guidance in reading this plot please refer to Figure \ref{fig:visual_aid}. \textbf{Left}: Unmodified WASP-121b run with mean molecular weight consistent with a molecular hydrogen-dominated atmosphere ($\mu_{m} = 2.35$). \textbf{Right}: Modified run with mean molecular weight $\mu_{m} = 1$. }
    \label{fig:los_terminatoravg_tavg_worotation_uhj}
\end{figure*}

In previous sections, we focused on how increasing irradiation affects the formation of dynamical features. As our sample transitions toward the ultra-hot Jupiter regime, the dissociation of H\textsubscript{2} and resulting changes in atmospheric composition are expected to also influence these processes. To explore this, Figure~\ref{fig:los_terminatoravg_tavg_worotation_uhj} compares our original WASP-121b simulation with a modified “mock ultra-hot Jupiter” version. The key difference is in the jet structure: while the original WASP-121b shows multiple alternating jets, the modified version produces a simpler pattern dominated by a strong central jet. This suggests that the modification pushes the circulation profile to resemble those of cooler hot Jupiters, in line with previous findings \citep{bell_cowan_2018,jones_2022}.

Physically, this can be understood through the framework proposed by \citet{showman_doppler_2013}, who argued that shortening radiative timescales can disrupt the formation of the smaller-scale dynamical structures that feed into jet formation. By changing the effective specific heat capacity in our modified simulation to mimic a smaller mean molecular weight, we effectively alter the radiative-dynamic balance. This acts to stabilize dynamical features and making it appear more zonal-jet-dominated-like. Checking the heat circulation efficiency (Figure~\ref{fig:perna_plot}) confirms that, although we are near the threshold where the trend of increasing irradiation correlating with decreased heat circulation efficiencies may break, our idealized approach of recreating a hot Jupiter–to–ultra-hot Jupiter transition improves the circulation efficiency.

Lastly, it is worth noting that the vertical extent of our models does not match perfectly across all simulations (Figure~\ref{fig:los_terminatorslice_tavg_worotation}), because the lowest pressure attainable on the dayside depends on the instellation. This known limitation of the THOR GCM \citep{deitrick_thor2.0_2020}, shared by other altitude-based GCMs, may slightly alter the upper atmospheric temperature and wind structures. However, this does not fundamentally affect our main conclusions regarding circulation trends and the influence of extreme stellar forcing.

\subsection{Opportunities for Future Work}
Our presented work aims to analyze parameter space-wide general patterns over a large sample at the cost of decreased complexity. A direct avenue of improvement would be to improve upon our "mock ultra-hot Jupiter" case by modeling their atmospheres rigorously and choosing a sample within the hotter ultra-hot Jupiter regime to further probe the effects of thermochemical processes on the atmospheres of these gas giant objects. Additional extension possibilities also exist, as there are a myriad of physical processes that could influence the global flow regime transition we have explored in our work. Some important ones would be to explore the effect of non-solar metallicities, the effect of chemical kinetics and/or clouds on the presented wave patterns.

\section{Conclusion}
\label{section:Conclusion}
We confirm the previously reported trend that increasing equilibrium temperature generally leads to a reduction in atmospheric heat circulation efficiency. Our results also provide evidence for a reversal in this trend, as suggested by \citet{bell_cowan_2018}, around $T_{\rm eq} \approx 2500$ K. By artificially altering the mean molecular weight in our WASP-121b simulations, we mimic an upper limit for the expected effects of reaching this ultra-hot Jupiter regime and demonstrate a corresponding improvement in circulation efficiency.

Additionally, our exploration of transitional flow regimes, as discussed by \citet{showman_doppler_2013}, suggests that these transitions may extend beyond what was initially proposed. \citet{showman_doppler_2013} identified the transition between jet-dominated and day-to-night flow within a three-planet sample (GJ 436b, HD 189733b, HD 209458b) spanning $T_{\mathrm{eq}}\ \approx\ 650~\text{–}~1500~K$. While we demonstrate the same flow transition as they propose, in their work, \citet{showman_doppler_2013} present HD 189733b as a transitionary case and HD 209458b is given as an example for a day-to-night-type flow with any hotter planet falling automatically into this category. In our findings, spanning a range of  $T_{\mathrm{eq}}\ \approx\ 1100~\text{–}~2400~K$,  HD 189733b and WASP-43b show zonal-jet-dominated flows, HD 209458b, WASP-31b, and WASP-17b occupy a transitional regime, and WASP-74b, WASP-19b, and WASP-121b exhibit day-to-night flow. We would like to highlight the fact that, while our findings are also model-dependent and are not exhaustive, they suggest that the flow regime transition is not a sharp cutoff at HD 209458b. Rather, it depends on multiple factors, such as the atmospheric composition, as demonstrated in our idealized case, and is better characterized by the coupling or decoupling of underlying wave structures.

We approach this problem through the application of our novel wave analysis method to a sample of eight planets spanning from warm Jupiters to an ultra-hot Jupiter-like regime. We identify a shift in the relative importance of smaller-scale eddy modes and their coupling to the mean flow. At lower equilibrium temperatures, both $\hat{k}=1$ and $\hat{k}=2$ modes contribute to sustaining equatorial superrotation. However, at higher irradiation levels, the $\hat{k}=2$ mode weakens, leaving a predominantly planetary-scale circulation. This change corresponds to a regime where day-to-night winds dominate, diminishing the dynamical influence of smaller-scale features. 

Our results support the theoretical frameworks put forth by \citet{showman_polvani_2011}, \citet{showman_doppler_2013} and the analysis performed by \citet{mendonca_2020}, by explicitly identifying the interaction between Rossby and Kelvin waves. Taken together, the extent of our parameter space demonstrates that the key dynamical turning point is the progressive decoupling of the small-scale $\hat{k} \approx 2$ modes from the dominant planetary-scale $\hat{k}=1$ mode.

Overall, these findings refine our understanding of how atmospheric circulation patterns evolve with increasing stellar insolation and provide a more comprehensive framework for interpreting the dynamical behavior of hot and ultra-hot Jupiter atmospheres. They also highlight the importance of capturing the interplay between large-scale modes and smaller-scale eddies and point the way toward more sophisticated modeling efforts. In doing so, this work can help guide future observations and modeling studies aimed at better understanding these extreme atmospheric regimes.

\bibliographystyle{aa.bst}
\bibliography{bib}

\begin{thebibliography}{84}
\expandafter\ifx\csname natexlab\endcsname\relax\def\natexlab#1{#1}\fi

\bibitem[{{Amundsen} {et~al.}(2016){Amundsen}, {Mayne}, {Baraffe}, {Manners}, {Tremblin}, {Drummond}, {Smith}, {Acreman}, \& {Homeier}}]{amundsen_hd209_2016}
{Amundsen}, D.~S., {Mayne}, N.~J., {Baraffe}, I., {et~al.} 2016, \aap, 595, A36

\bibitem[{{Anderson} {et~al.}(2011){Anderson}, {Collier Cameron}, {Gillon}, {Hellier}, {Jehin}, {Lendl}, {Maxted}, {Queloz}, {Smalley}, {Smith}, {Triaud}, {West}, {Pepe}, {S{\'e}gransan}, {Udry}, {Basri}, {Bouchy}, \& {Vidal-Madjar}}]{anderson2011}
{Anderson}, D.~R., {Collier Cameron}, A., {Gillon}, M., {et~al.} 2011, \aap, 528, A97

\bibitem[{{Arcangeli} {et~al.}(2018){Arcangeli}, {D{\'e}sert}, {Line}, {Bean}, {Parmentier}, {Stevenson}, {Kreidberg}, {Fortney}, {Mansfield}, \& {Showman}}]{arcangeli_2018}
{Arcangeli}, J., {D{\'e}sert}, J.-M., {Line}, M.~R., {et~al.} 2018, \apjl, 855, L30

\bibitem[{{Bell} \& {Cowan}(2018)}]{bell_cowan_2018}
{Bell}, T.~J. \& {Cowan}, N.~B. 2018, \apjl, 857, L20

\bibitem[{{Boyajian} {et~al.}(2015){Boyajian}, {von Braun}, {Feiden}, {Huber}, {Basu}, {Demory}, {Fischer}, {Schaefer}, {Mann}, {White}, {Maestro}, {Brewer}, {Lamell}, {Spada}, {L{\'o}pez-Morales}, {Ireland}, {Farrington}, {van Belle}, {Kane}, {Jones}, {Behr}, {Tenenbaum}, {Ciardi}, {McAlister}, {Ridgway}, {Goldfinger}, {Turner}, \& {Sturmann}}]{boyajian2015}
{Boyajian}, T.~S., {von Braun}, K., {Feiden}, G.~A., {et~al.} 2015, \apj, 800, 51

\bibitem[{{Cartes} {et~al.}(2020){Cartes}, {Zuleta}, {L{\'o}pez}, \& {Rojo}}]{cartes-zuleta2020}
{Cartes}, Z.~E., {Zuleta}, J., {L{\'o}pez}, R., \& {Rojo}, P. 2020, \aap, 642, A29

\bibitem[{{Chandrasekhar}(1935)}]{1935MNRAS..96...21C}
{Chandrasekhar}, S. 1935, \mnras, 96, 21

\bibitem[{{Cooper} \& {Showman}(2005)}]{showman_cooper_2005}
{Cooper}, C.~S. \& {Showman}, A.~P. 2005, \apjl, 629, L45

\bibitem[{{Daubechies} \& {Bates}(1993)}]{ten_lectures_on_wavelets_1993}
{Daubechies}, I. \& {Bates}, B.~J. 1993, Acoustical Society of America Journal, 93, 1671

\bibitem[{{Dawson} \& {Johnson}(2018)}]{dawson_2018}
{Dawson}, R.~I. \& {Johnson}, J.~A. 2018, \araa, 56, 175

\bibitem[{Deitrick {et~al.}(2022)Deitrick, Heng, Schroffenegger, Kitzmann, Grimm, Malik, Mendon{\c c}a, \& Morris}]{deitrick_thor_2022}
Deitrick, R., Heng, K., Schroffenegger, U., {et~al.} 2022, \mnras, 512, 3759

\bibitem[{Deitrick {et~al.}(2020)Deitrick, Mendon{\c c}a, Schroffenegger, Grimm, Tsai, \& Heng}]{deitrick_thor2.0_2020}
Deitrick, R., Mendon{\c c}a, J.~M., Schroffenegger, U., {et~al.} 2020, \apjs, 248, 30

\bibitem[{{Delrez} {et~al.}(2016){Delrez}, {Santerne}, {Almenara}, {Anderson}, {Collier Cameron}, {D{\'\i}az}, {Gillon}, {Hellier}, {Holman}, {Jehin}, {Lendl}, {Maxted}, {Neveu-VanMalle}, {Pepe}, {Pollacco}, {Queloz}, {S{\'e}gransan}, {Smalley}, {Smith}, {Southworth}, {Triaud}, {Udry}, {Van Grootel}, \& {West}}]{delrez2016}
{Delrez}, L., {Santerne}, A., {Almenara}, J.~M., {et~al.} 2016, \mnras, 458, 4025

\bibitem[{{Dobbs-Dixon} \& {Agol}(2013)}]{dobbs-dixon_agol_2013}
{Dobbs-Dixon}, I. \& {Agol}, E. 2013, \mnras, 435, 3159

\bibitem[{Drummond {et~al.}(2018)Drummond, Mayne, Manners, Baraffe, Goyal, Tremblin, Sing, \& Kohary}]{drummond_3d_2018}
Drummond, B., Mayne, N.~J., Manners, J., {et~al.} 2018, The Astrophysical Journal, 869, 28

\bibitem[{Emanuel(1994)}]{emanuel1994atmospheric}
Emanuel, K. 1994, Atmospheric Convection (Oxford University Press)

\bibitem[{Farge(1992)}]{farge_wavelet_1992}
Farge, M. 1992, Annual Review of Fluid Mechanics, 24, 395, publisher: Annual Reviews

\bibitem[{{Flowers} {et~al.}(2019){Flowers}, {Brogi}, {Rauscher}, {Kempton}, \& {Chiavassa}}]{flowers_hd189_2019}
{Flowers}, E., {Brogi}, M., {Rauscher}, E., {Kempton}, E. M.~R., \& {Chiavassa}, A. 2019, \aj, 157, 209

\bibitem[{{Freedman} {et~al.}(2014){Freedman}, {Lustig-Yaeger}, {Fortney}, {Lupu}, {Marley}, \& {Lodders}}]{freedman_2014}
{Freedman}, R.~S., {Lustig-Yaeger}, J., {Fortney}, J.~J., {et~al.} 2014, \apjs, 214, 25

\bibitem[{Gabor(1946)}]{gabor_1945}
Gabor, D. 1946, Journal of the Institution of Electrical Engineers - Part III: Radio and Communication Engineering, 93, 445

\bibitem[{Gill(1982)}]{gill_1982}
Gill, A. 1982, Atmosphere-Ocean Dynamics, Atmosphere-Ocean Dynamics No. v. 30 (Elsevier Science)

\bibitem[{Gill(1980)}]{gill_simple_1980}
Gill, A.~E. 1980, Quarterly Journal of the Royal Meteorological Society, 106, 447

\bibitem[{{Guillot} {et~al.}(1996){Guillot}, {Burrows}, {Hubbard}, {Lunine}, \& {Saumon}}]{guillot_1996}
{Guillot}, T., {Burrows}, A., {Hubbard}, W.~B., {Lunine}, J.~I., \& {Saumon}, D. 1996, \apjl, 459, L35

\bibitem[{Hammond \& Abbot(2022)}]{hammond_numerical_2022}
Hammond, M. \& Abbot, D.~S. 2022, \mnras, 511, 2313

\bibitem[{{Hammond} \& {Lewis}(2021)}]{hammond_lewis_2021}
{Hammond}, M. \& {Lewis}, N.~T. 2021, Proceedings of the National Academy of Science, 118, e2022705118

\bibitem[{{Hellier} {et~al.}(2011){Hellier}, {Anderson}, {Collier Cameron}, {Gillon}, {Jehin}, {Lendl}, {Maxted}, {Queloz}, {Smalley}, {Smith}, {Triaud}, \& {West}}]{hellier2011}
{Hellier}, C., {Anderson}, D.~R., {Collier Cameron}, A., {et~al.} 2011, \aap, 534, A20

\bibitem[{Heng \& Kitzmann(2017)}]{heng_analytical_2017}
Heng, K. \& Kitzmann, D. 2017, \apjs, 232, 20, aDS Bibcode: 2017ApJS..232...20H

\bibitem[{Heng {et~al.}(2018)Heng, Malik, \& Kitzmann}]{heng_analytical_2018}
Heng, K., Malik, M., \& Kitzmann, D. 2018, \apjs, 237, 29, aDS Bibcode: 2018ApJS..237...29H

\bibitem[{Heng {et~al.}(2011)Heng, Menou, \& Phillipps}]{heng_2011a}
Heng, K., Menou, K., \& Phillipps, P.~J. 2011, \mnras, 413, 2380

\bibitem[{{Heng} \& {Showman}(2015)}]{heng_showman_2015}
{Heng}, K. \& {Showman}, A.~P. 2015, Annual Review of Earth and Planetary Sciences, 43, 509

\bibitem[{Heng \& Workman(2014)}]{heng_analytical_2014}
Heng, K. \& Workman, J. 2014, \apjs, 213, 27, aDS Bibcode: 2014ApJS..213...27H

\bibitem[{{Holton}(1992)}]{holton_1992}
{Holton}, J.~R. 1992, {An introduction to dynamic meteorology}

\bibitem[{Jablonowski \& Williamson(2011)}]{Jablonowski2011}
Jablonowski, C. \& Williamson, D.~L. 2011, in Numerical Techniques for Global Atmospheric Models, ed. P.~H. Lauritzen, C.~Jablonowski, M.~A. Taylor, \& R.~Ramachandran (Berlin: Springer), 381--493

\bibitem[{{Jones} {et~al.}(2022){Jones}, {Morris}, {Demory}, {Heng}, {Hooton}, {Billot}, {Ehrenreich}, {Hoyer}, {Simon}, {Lendl}, {Demangeon}, {Sousa}, {Bonfanti}, {Wilson}, {Salmon}, {Csizmadia}, {Parviainen}, {Bruno}, {Alibert}, {Alonso}, {Anglada}, {B{\'a}rczy}, {Barrado}, {Barros}, {Baumjohann}, {Beck}, {Beck}, {Benz}, {Bonfils}, {Brandeker}, {Broeg}, {Cabrera}, {Charnoz}, {Collier Cameron}, {Davies}, {Deleuil}, {Deline}, {Delrez}, {Erikson}, {Fortier}, {Fossati}, {Fridlund}, {Gandolfi}, {Gillon}, {G{\"u}del}, {Isaak}, {Kiss}, {Laskar}, {Lecavelier des Etangs}, {Lovis}, {Magrin}, {Maxted}, {Nascimbeni}, {Olofsson}, {Ottensamer}, {Pagano}, {Pall{\'e}}, {Peter}, {Piotto}, {Pollacco}, {Queloz}, {Ragazzoni}, {Rando}, {Ratti}, {Rauer}, {Reimers}, {Ribas}, {Santos}, {Scandariato}, {S{\'e}gransan}, {Smith}, {Steller}, {Szab{\'o}}, {Thomas}, {Udry}, {Van Grootel}, {Walter}, {Walton}, \& {Wang Jungo}}]{jones_2022}
{Jones}, K., {Morris}, B.~M., {Demory}, B.~O., {et~al.} 2022, \aap, 666, A118

\bibitem[{Kataria {et~al.}(2016)Kataria, Sing, Lewis, Visscher, Showman, Fortney, \& Marley}]{kataria_atmospheric_2016}
Kataria, T., Sing, D.~K., Lewis, N.~K., {et~al.} 2016, The Astrophysical Journal, 821, 9

\bibitem[{{Knutson} {et~al.}(2012){Knutson}, {Lewis}, {Fortney}, {Burrows}, {Showman}, {Cowan}, {Agol}, {Aigrain}, {Charbonneau}, {Deming}, {D{\'e}sert}, {Henry}, {Langton}, \& {Laughlin}}]{knutson_hd189_2012}
{Knutson}, H.~A., {Lewis}, N., {Fortney}, J.~J., {et~al.} 2012, \apj, 754, 22

\bibitem[{{Komacek} \& {Showman}(2016)}]{komacek_showman_2016}
{Komacek}, T.~D. \& {Showman}, A.~P. 2016, \apj, 821, 16

\bibitem[{Lee {et~al.}(2021)Lee, Parmentier, Hammond, Grimm, Kitzmann, Tan, Tsai, \& Pierrehumbert}]{lee_simulating_2021}
Lee, E. K.~H., Parmentier, V., Hammond, M., {et~al.} 2021, \mnras, 506, 2695

\bibitem[{{Lee} {et~al.}(2022){Lee}, {Prinoth}, {Kitzmann}, {Tsai}, {Hoeijmakers}, {Borsato}, \& {Heng}}]{lee_w121b_2022}
{Lee}, E. K.~H., {Prinoth}, B., {Kitzmann}, D., {et~al.} 2022, \mnras, 517, 240

\bibitem[{Lee {et~al.}(2023)Lee, Gommers, Wohlfahrt, Wasilewski, O'Leary, {Holger}, Sauvé, Millman, Agrawal, Clauss, Pelt, Oliveira, Yu, Brett, Pelletier, {SylvainLan}, Tricoli, Choudhary, Solak, {asnt}, Smith, {0-tree}, Goldberg, Goertzen, Laszuk, {ElConno}, Ocansey, Antonello, Mandula, \& {jakirkham}}]{lee_pywaveletspywt_2023}
Lee, G., Gommers, R., Wohlfahrt, K., {et~al.} 2023, {PyWavelets}/pywt: v1.5.0

\bibitem[{{Lewis} \& {Hammond}(2022)}]{lewis_hammond_2022}
{Lewis}, N.~T. \& {Hammond}, M. 2022, \apj, 941, 171

\bibitem[{{Li} \& {Goodman}(2010)}]{li_goodman_2010}
{Li}, J. \& {Goodman}, J. 2010, \apj, 725, 1146

\bibitem[{Malik {et~al.}(2017)Malik, Grosheintz, Mendonça, Grimm, Lavie, Kitzmann, Tsai, Burrows, Kreidberg, Bedell, Bean, Stevenson, \& Heng}]{malik_helios_2017}
Malik, M., Grosheintz, L., Mendonça, J.~M., {et~al.} 2017, \aj, 153, 56, aDS Bibcode: 2017AJ....153...56M

\bibitem[{Malik {et~al.}(2019)Malik, Kitzmann, Mendonça, Grimm, Marleau, Linder, Tsai, \& Heng}]{malik_self-luminous_2019}
Malik, M., Kitzmann, D., Mendonça, J.~M., {et~al.} 2019, \aj, 157, 170, aDS Bibcode: 2019AJ....157..170M

\bibitem[{Matsuno(1966)}]{matsuno_quasi-geostrophic_1966}
Matsuno, T. 1966, Journal of the Meteorological Society of Japan, 44, 25, aDS Bibcode: 1966JMeSJ..44...25M

\bibitem[{{Mayne} {et~al.}(2017){Mayne}, {Debras}, {Baraffe}, {Thuburn}, {Amundsen}, {Acreman}, {Smith}, {Browning}, {Manners}, \& {Wood}}]{mayne_2017}
{Mayne}, N.~J., {Debras}, F., {Baraffe}, I., {et~al.} 2017, \aap, 604, A79

\bibitem[{Mendon{\c c}a {et~al.}(2018{\natexlab{a}})Mendon{\c c}a, Malik, Demory, \& Heng}]{mendonca_w43b_2018a}
Mendon{\c c}a, J.~M., Malik, M., Demory, B.-O., \& Heng, K. 2018{\natexlab{a}}, \aj, 155, 150

\bibitem[{Mendon{\c c}a {et~al.}(2018{\natexlab{b}})Mendon{\c c}a, Tsai, Malik, Grimm, \& Heng}]{mendonca_w43b_2018b}
Mendon{\c c}a, J.~M., Tsai, S.-m., Malik, M., Grimm, S.~L., \& Heng, K. 2018{\natexlab{b}}, The Astrophysical Journal, 869, 107

\bibitem[{{Mendon{\c{c}}a}(2020)}]{mendonca_2020}
{Mendon{\c{c}}a}, J.~M. 2020, \mnras, 491, 1456

\bibitem[{{Mendon{\c{c}}a} {et~al.}(2016){Mendon{\c{c}}a}, {Grimm}, {Grosheintz}, \& {Heng}}]{mendonca_thor_2016}
{Mendon{\c{c}}a}, J.~M., {Grimm}, S.~L., {Grosheintz}, L., \& {Heng}, K. 2016, \apj, 829, 115

\bibitem[{{Menou} {et~al.}(2003){Menou}, {Cho}, {Seager}, \& {Hansen}}]{menou_2003}
{Menou}, K., {Cho}, J. Y.~K., {Seager}, S., \& {Hansen}, B. M.~S. 2003, \apjl, 587, L113

\bibitem[{Moca {et~al.}(2021)Moca, Bârzan, Nagy-Dăbâcan, \& Mureșan}]{moca_time-frequency_2021}
Moca, V.~V., Bârzan, H., Nagy-Dăbâcan, A., \& Mureșan, R.~C. 2021, Nature Communications, 12, 337, publisher: Nature Publishing Group

\bibitem[{Morlet {et~al.}(1982)Morlet, Arens, Fourgeau, \& Giard}]{morlet_1982}
Morlet, J., Arens, G., Fourgeau, E., \& Giard, D. 1982, Geophysics, 47, 222

\bibitem[{Noti {et~al.}(2023)Noti, Lee, Deitrick, \& Hammond}]{noti_examining_2023}
Noti, P.~A., Lee, E. K.~H., Deitrick, R., \& Hammond, M. 2023, Examining {NHD} vs {QHD} in the {GCM} {THOR} with non-grey radiative transfer for the hot {Jupiter} regime

\bibitem[{{Parmentier} \& {Guillot}(2014)}]{parmentier_picket_fence_2014}
{Parmentier}, V. \& {Guillot}, T. 2014, \aap, 562, A133

\bibitem[{{Parmentier} {et~al.}(2015){Parmentier}, {Guillot}, {Fortney}, \& {Marley}}]{parmentier_picket_fence_2015}
{Parmentier}, V., {Guillot}, T., {Fortney}, J.~J., \& {Marley}, M.~S. 2015, \aap, 574, A35

\bibitem[{{Parmentier} {et~al.}(2018){Parmentier}, {Line}, {Bean}, {Mansfield}, {Kreidberg}, {Lupu}, {Visscher}, {D{\'e}sert}, {Fortney}, {Deleuil}, {Arcangeli}, {Showman}, \& {Marley}}]{parmentier_w121b_2018}
{Parmentier}, V., {Line}, M.~R., {Bean}, J.~L., {et~al.} 2018, \aap, 617, A110

\bibitem[{{Perez-Becker} \& {Showman}(2013)}]{perez_becker_showman_2013}
{Perez-Becker}, D. \& {Showman}, A.~P. 2013, \apj, 776, 134

\bibitem[{Perna {et~al.}(2012)Perna, Heng, \& Pont}]{perna_effects_2012}
Perna, R., Heng, K., \& Pont, F. 2012, The Astrophysical Journal, 751, 59

\bibitem[{{Perna} {et~al.}(2010){Perna}, {Menou}, \& {Rauscher}}]{perna_2010}
{Perna}, R., {Menou}, K., \& {Rauscher}, E. 2010, \apj, 719, 1421

\bibitem[{Showman {et~al.}(2013)Showman, Fortney, Lewis, \& Shabram}]{showman_doppler_2013}
Showman, A.~P., Fortney, J.~J., Lewis, N.~K., \& Shabram, M. 2013, The Astrophysical Journal, 762, 24

\bibitem[{{Showman} {et~al.}(2009){Showman}, {Fortney}, {Lian}, {Marley}, {Freedman}, {Knutson}, \& {Charbonneau}}]{showman_hd189_2009}
{Showman}, A.~P., {Fortney}, J.~J., {Lian}, Y., {et~al.} 2009, \apj, 699, 564

\bibitem[{{Showman} \& {Guillot}(2002)}]{showman_guillot_2002}
{Showman}, A.~P. \& {Guillot}, T. 2002, \aap, 385, 166

\bibitem[{{Showman} {et~al.}(2015){Showman}, {Lewis}, \& {Fortney}}]{showman_lewis_fortney_2015}
{Showman}, A.~P., {Lewis}, N.~K., \& {Fortney}, J.~J. 2015, \apj, 801, 95

\bibitem[{{Showman} \& {Polvani}(2011)}]{showman_polvani_2011}
{Showman}, A.~P. \& {Polvani}, L.~M. 2011, \apj, 738, 71

\bibitem[{{Showman} {et~al.}(2020){Showman}, {Tan}, \& {Parmentier}}]{showman_tan_parmentier_2020}
{Showman}, A.~P., {Tan}, X., \& {Parmentier}, V. 2020, \ssr, 216, 139

\bibitem[{{Skinner} \& {Cho}(2021)}]{skinner_cho_2021}
{Skinner}, J.~W. \& {Cho}, J.~Y.~K. 2021, \mnras, 504, 5172

\bibitem[{{Skinner} \& {Cho}(2022)}]{skinner_cho_2022}
{Skinner}, J.~W. \& {Cho}, J.~Y.~K. 2022, \mnras, 511, 3584

\bibitem[{{Skinner} \& {Cho}(2025)}]{skinner_cho_2025}
{Skinner}, J.~W. \& {Cho}, J.~Y.~K. 2025, \apj, 982, 64

\bibitem[{{Staniforth} \& {Thuburn}(2012)}]{staniforth_thuburn_2012}
{Staniforth}, A. \& {Thuburn}, J. 2012, Quarterly Journal of the Royal Meteorological Society, 138, 1

\bibitem[{{Stassun} {et~al.}(2017){Stassun}, {Corsaro}, {Pepper}, \& {Gaudi}}]{stassun2017}
{Stassun}, K.~G., {Corsaro}, E., {Pepper}, J.~A., \& {Gaudi}, B.~S. 2017, \aj, 153, 136

\bibitem[{{Steinrueck} {et~al.}(2019){Steinrueck}, {Parmentier}, {Showman}, {Lothringer}, \& {Lupu}}]{steinrueck_hd189_2019}
{Steinrueck}, M.~E., {Parmentier}, V., {Showman}, A.~P., {Lothringer}, J.~D., \& {Lupu}, R.~E. 2019, \apj, 880, 14

\bibitem[{Tan \& Komacek(2019)}]{tan_atmospheric_2019}
Tan, X. \& Komacek, T.~D. 2019, The Astrophysical Journal, 886, 26, arXiv:1910.01622 [astro-ph]

\bibitem[{Tan \& Showman(2021)}]{tan_atmospheric_2021}
Tan, X. \& Showman, A.~P. 2021, \mnras, 502, 2198, aDS Bibcode: 2021MNRAS.502.2198T

\bibitem[{Tomita \& Satoh(2004)}]{tomita_new_2004}
Tomita, H. \& Satoh, M. 2004, Fluid Dynamics Research, 34, 357

\bibitem[{{Tomita} {et~al.}(2002){Tomita}, {Satoh}, \& {Goto}}]{tomita_satoh_2002}
{Tomita}, H., {Satoh}, M., \& {Goto}, K. 2002, Journal of Computational Physics, 183, 307

\bibitem[{Tomita {et~al.}(2001)Tomita, Tsugawa, Satoh, \& Goto}]{tomita_shallow_2001}
Tomita, H., Tsugawa, M., Satoh, M., \& Goto, K. 2001, Journal of Computational Physics, 174, 579

\bibitem[{Torrence \& Compo(1998)}]{torrence_practical_1998}
Torrence, C. \& Compo, G.~P. 1998, Bulletin of the American Meteorological Society, 79, 61

\bibitem[{{Torres} {et~al.}(2008){Torres}, {Winn}, \& {Holman}}]{torres2008}
{Torres}, G., {Winn}, J.~N., \& {Holman}, M.~J. 2008, \apj, 677, 1324

\bibitem[{{Tsai} {et~al.}(2014){Tsai}, {Dobbs-Dixon}, \& {Gu}}]{tsai_2014}
{Tsai}, S.-M., {Dobbs-Dixon}, I., \& {Gu}, P.-G. 2014, \apj, 793, 141

\bibitem[{{Wheeler} \& {Kiladis}(1999)}]{wheeler_convectively_1999}
{Wheeler}, M. \& {Kiladis}, G.~N. 1999, Journal of the Atmospheric Sciences, 56, 374

\bibitem[{{Wicker} \& {Skamarock}(2002)}]{wicker_skamarock_2002}
{Wicker}, L.~J. \& {Skamarock}, W.~C. 2002, Monthly Weather Review, 130, 2088

\bibitem[{{Wong} {et~al.}(2016){Wong}, {Knutson}, {Kataria}, {Lewis}, {Burrows}, {Fortney}, {Schwartz}, {Shporer}, {Agol}, {Cowan}, {Deming}, {D{\'e}sert}, {Fulton}, {Howard}, {Langton}, {Laughlin}, {Showman}, \& {Todorov}}]{wong_wasp19b_2016}
{Wong}, I., {Knutson}, H.~A., {Kataria}, T., {et~al.} 2016, \apj, 823, 122

\bibitem[{{Zellem} {et~al.}(2014){Zellem}, {Griffith}, {Lewis}, {Swain}, \& {Knutson}}]{zellem_hd209_2014}
{Zellem}, R., {Griffith}, C.~A., {Lewis}, N.~K., {Swain}, M.~R., \& {Knutson}, H.~A. 2014, in AAS/Division for Planetary Sciences Meeting Abstracts, Vol.~46, AAS/Division for Planetary Sciences Meeting Abstracts \#46, 104.04

\end{thebibliography}

\newpage
\begin{appendix}
\section{Additional Wave Analysis Results}
\label{appendix: Additional Wave Analysis Results}
\begin{figure*}
    \centering
    \caption{Spatial distribution of wavelet coefficients of the large-scale mode $\hat{k} = 0.3$ at the 0.1\,bar atmospheric layer. The shown results are time-averaged over the last 100 days of the analysis period.}
    \label{fig:wavelet_large_scale}
    \begin{subfigure}{\textwidth}
        \centering
        \caption{HD~189733b}
        \includegraphics[width=\columnwidth]{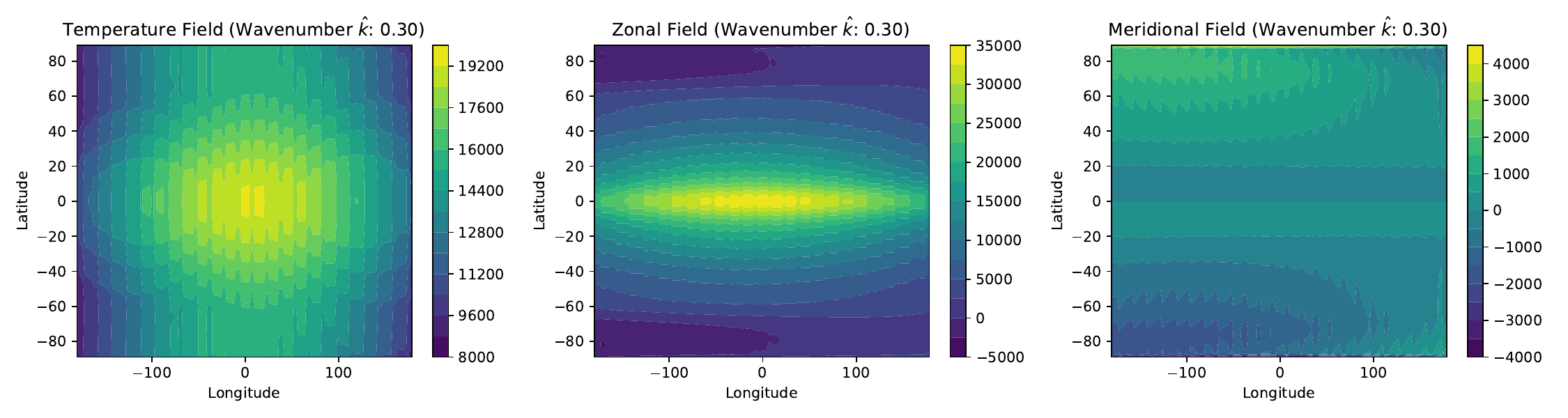}
    \end{subfigure}
    \vspace{0.25cm}
    \begin{subfigure}{\textwidth}
        \centering
        \caption{HD~209458b}
        \includegraphics[width=\columnwidth]{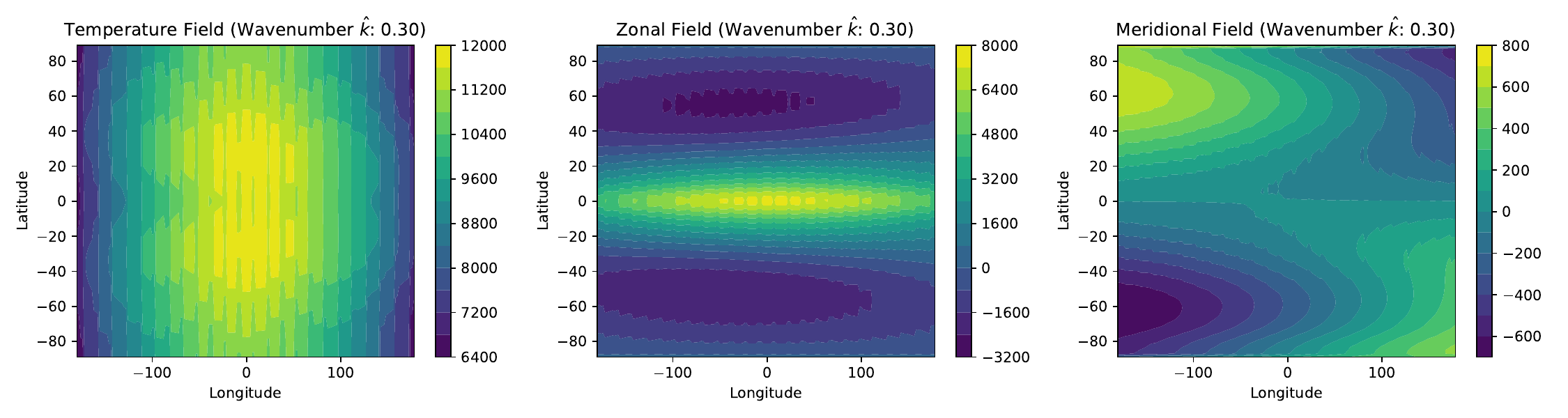}
    \end{subfigure}
    \vspace{0.25cm}
    \begin{subfigure}{\textwidth}
        \centering
        \caption{WASP-121b}
        \includegraphics[width=\columnwidth]{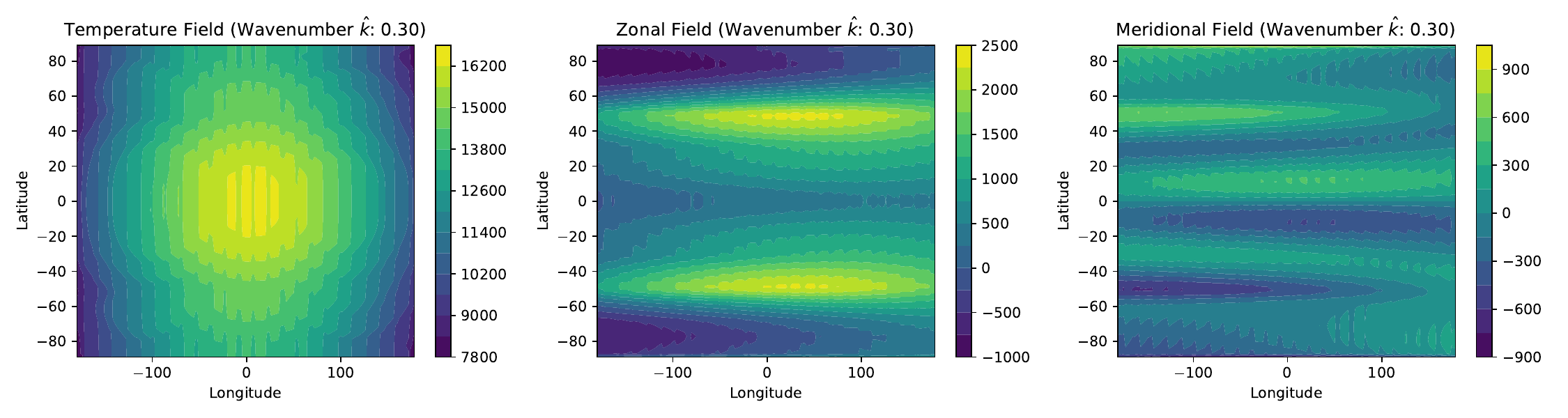}
    \end{subfigure}
\end{figure*}

\begin{figure*}
    \centering
    \caption{\textbf{Left panel:} Snapshot of the GCM solutions for  HD 189733b, HD~209458b and WASP-121b at 300 days after the full simulation time ($\approx$ 4000-5000 days), showing temperature and wind fields at the approximate 0.1~bar atmospheric layer. \textbf{Center \& Right:} Wavelet scalograms of the zonal wind field at the equator and mid-latitudes, respectively.}
    \label{fig:wavelet_last_500}
    \begin{subfigure}{\textwidth}
        \centering
        \caption{HD~189733b}
        \includegraphics[width=\columnwidth]{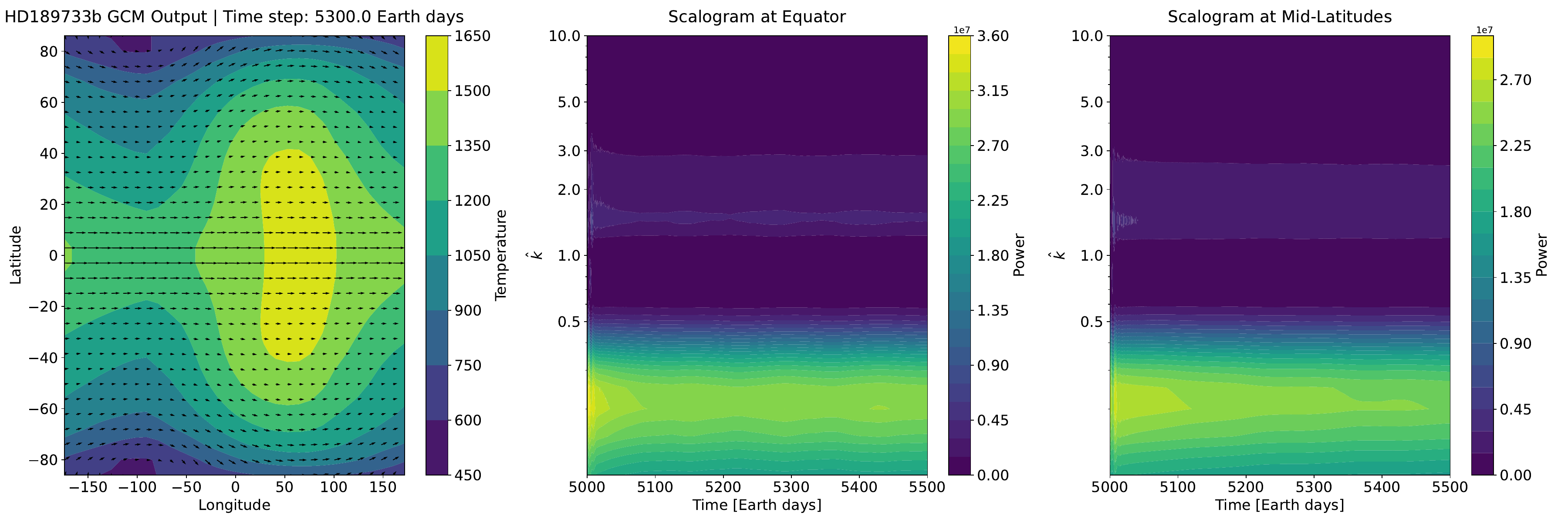}
    \end{subfigure}
    \vspace{0.25cm}
    \begin{subfigure}{\textwidth}
        \centering
        \caption{HD~209458b}
        \includegraphics[width=\columnwidth]{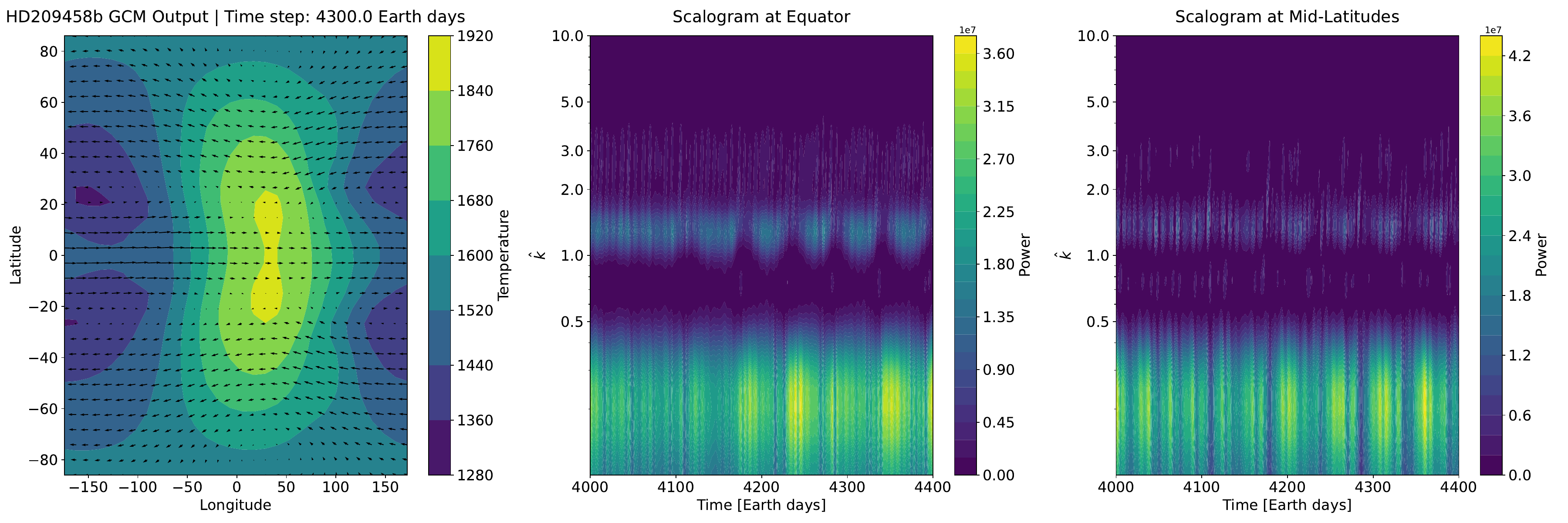}
    \end{subfigure}
    \vspace{0.25cm}
    \begin{subfigure}{\textwidth}
        \centering
        \caption{WASP-121b}
        \includegraphics[width=\columnwidth]{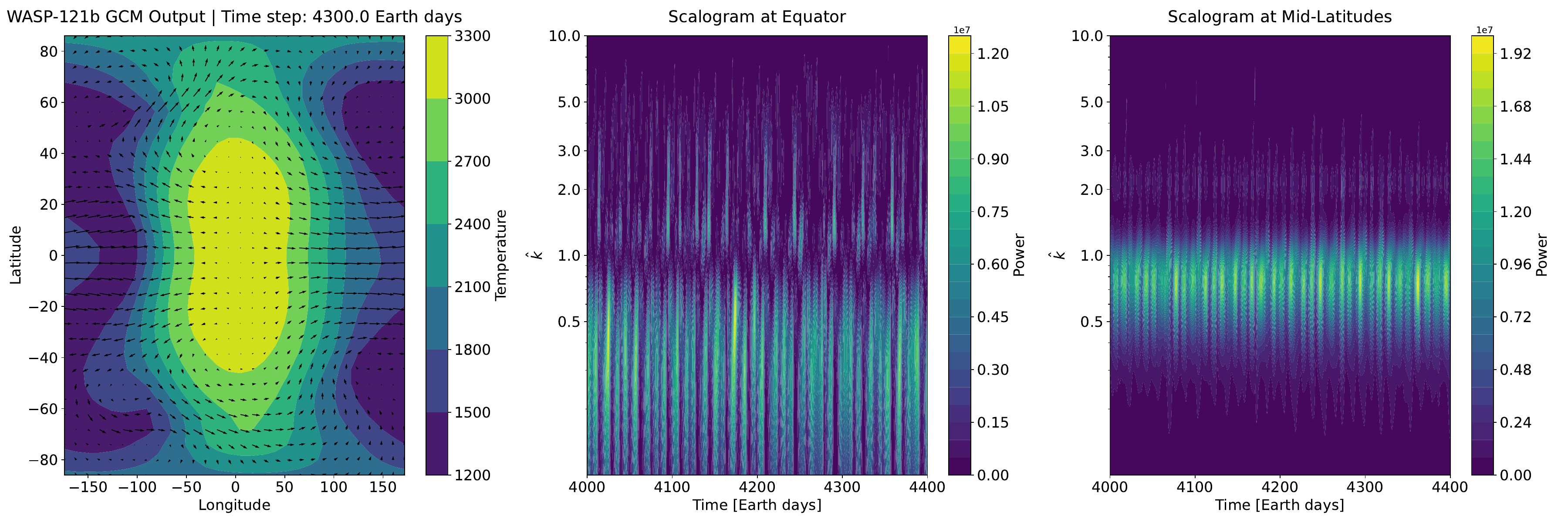}
    \end{subfigure}
\end{figure*}

\subsection{Large-Scale Wave Modes}
In Figure~\ref{fig:wavelet_large_scale}, we show the spatial distribution of the wavelet coefficients for the temperature, zonal wind, and meridional wind fields on the $\approx 0.1$\,bar atmospheric layer for HD~189733b, HD~209458b, and WASP-121b. We focus on modes $\hat{k}=0.3$ as a proxy for the large-scale features discussed in Section~\ref{section: Applying Wavelet Analysis to GCM outputs}. 

For HD~189733b (Figure~\ref{fig:wavelet_large_scale}a), a dominant equatorial structure in the zonal wind field appears along all longitudes, representing the zonal-jet-dominated mean flow. In HD~209458b (Figure~\ref{fig:wavelet_large_scale}b), the large-scale equatorial mean flow is also present, and we begin to see symmetrical contributions from the mid-latitude regions. By contrast, WASP-121b (Figure~\ref{fig:wavelet_large_scale}c) does not display a robust equatorial zonal wind feature and instead exhibits more symmetric structures in the mid-latitudes. An additional effect that is present in all of these plots is that the wavelet coefficients appear to be organized in vertical bands, looking as if the underlying field has not been sampled on a fine enough grid. What we are observing here is again the Heisenberg-Gabor uncertainty, which becomes more pronounced for lower wavenumbers $\hat{k}$, i.e. larger scales.

\subsection{Wavelet Analysis of Steady-State Solutions}
For completeness, Figure \ref{fig:wavelet_last_500} shows scalograms for the final 500 days of the total simulation time in our example cases. Focusing on HD 189733b, on the left panel, the equatorial superrotation feature has gotten broader in latitude and is the only clear dynamical feature. The hotspot offset is also more pronounced than in Figure \ref{fig:hd189b_wavelet_decomp}. Both the equatorial and mid-latitude scalograms look nearly identical, featuring a broad, large-scale mode ($\hat{k} \approx 0.2\text{–}0.3$) and a smaller-scale mode ($\hat{k} = 2$). This observation aligns with physical expectations, as the large-scale mode corresponds to the steady-state zonal jet. The similarity between the equatorial and mid-latitude scalograms indicates that the dynamical features are strongly coupled in both regions. Furthermore, the $\hat{k} = 2$ mode persists as a response continuously triggered by stellar irradiation and planetary rotation, transferring momentum up-gradient. Both modes' temporal frequencies have lengthened enough to appear nearly constant, supporting our interpretation that this represents a steady-state standing wave.

HD 209458b represents our transitional case. From the middle-left panel, it is evident that it closely resembles its early stages presented in \ref{fig:hd209b_wavelet_decomp}. The equatorial jet feature is less pronounced and slightly narrower, mid- and high-latitude westward flows remain, and the hotspot is less prominent, indicating that heat has been transported to some extent around the planet. The scalograms are qualitatively very similar to those of the HD 189733b case. Again, we observe a large-scale mode ($\hat{k} \approx 0.2\text{-}0.3$) and a smaller-scale mode ($\hat{k} = 1.5\text{-}2$). In addition, there seems to be a weak contribution from a planetary-scale wave mode at $\hat{k} \approx 0.8$. While all inspected wave modes remain constant in power throughout the analysis period, apart from intermittent spikes in power in the small-scale mode, suggesting a steady state, these modes appear to oscillate with a high temporal frequency, in contrast to the HD 189733b case. This implies that, although the circulation is still dynamically efficient enough to form the large-scale superrotating feature, the increased instellation disrupts its progression into a fully zonal-jet-dominated regime.

WASP-121b is the hottest planet in our sample and represents our day-to-night flow case. Although the left panel shows a flow regime that is similar to what is shown in Figure \ref{fig:wasp121b_wavelet_decomp}, the scalograms present a significantly different picture. At the equator, two very broad signals span almost the entire wavenumber range: one at $\hat{k} = 0.1 - 0.8$ and another centered around $\hat{k} \approx 1.5$, with intermittent spikes extending to $\hat{k} \approx 4 - 5$. Combining this observation with the snapshot on the left panel, we conclude that no dynamical wave features can form apart from the day-to-night flow, which arises due to the thermal gradient between the day- and nightsides. The existence of these small-scale features hints that wave responses do get excited in the GCM solutions but do not couple to the main flow, as the short radiative timescales in this region disrupt the formation of any dynamical structures. At mid-latitudes, we identify two wave modes: one planetary-scale mode with $\hat{k} \approx 1$ and one small-scale mode with $\hat{k} \approx 2$. Although no large-scale feature is identified, the presence of these modes suggests that the mid-latitude region may be cool enough, i.e., the radiative timescale is sufficiently long, for some dynamical structures to form.

\end{appendix}
\end{document}